\documentclass{jfm}
\usepackage{graphicx}
\usepackage{natbib}
\usepackage{epstopdf, epsfig}
\usepackage{array}
\usepackage{tabularx}
\usepackage{enumerate}
\usepackage{amsmath}
\usepackage{slashed}
\usepackage{wrapfig}
\usepackage[dvipsnames]{xcolor}
\usepackage{comment}
\usepackage{multirow}

\shorttitle{Capsules in a T-junction}
\shortauthor{E. H\"aner, M. Heil and A. Juel}

\title{Deformation and sorting of capsules in a T-junction}

\author{Edgar H\"aner\aff{1}, Matthias Heil\aff{2}
 \and Anne Juel\aff{1} \corresp{\email{anne.juel@manchester.ac.uk}},}

\affiliation{\aff{1} MCND and School of Physics \& Astronomy, University of Manchester, Oxford Road, Manchester M13 9PL, UK\\
\aff{2} School of Mathematics and MCND, University of Mathematics, Oxford Road, Manchester M13 9PL, UK}

\begin{document}
	
\maketitle
\begin{abstract}
We study experimentally the motion and deformation of individual capsules transported by a constant volume-flux flow of low Reynolds number, through the T-junction of a channel with rectangular cross-section. We use millimetric ovalbumin-alginate capsules which we manufacture and characterise independently of the flow experiment. Centred capsules travel at constant velocity down the straight channel leading to the T-junction where they decelerate and expand in the spanwise direction before turning into one of the two identical daughter channels. There, non-inertial lift forces act to re-centre them and relax their shape until they reach a steady state of propagation. We find that the dynamics of fixed-size capsules within our channel geometry are governed by a capillary number $Ca$ defined as the ratio of viscous shear forces to elastic restoring forces. We quantify the elastic forces by statically compressing the capsule to 50\% of its initial diameter between parallel plates rather than by the Young's modulus of the encapsulating membrane, in order to account for different membrane thickness, pre-inflation and non-linear elastic deformation. We show that the maximum extension in the T-junction of capsules of different stiffness collapses onto a master curve in $Ca$. Thus, it provides a sensitive measure of the relative stiffness of capsules at constant flow rate, particularly for softer capsules. We also find that the T-junction can sort fixed-size capsules according to their stiffness because the position in the T-junction from which capsules are entrained into the daughter channel depends uniquely on $Ca$. We demonstrate that a T-junction can be used as a sorting device by enhancing this initial capsule separation through a diffuser.
\end{abstract}


\section{Introduction}

 The encapsulation of liquid droplets by an elastic membrane is an effective means of isolating active substances such as aromas, flavours, drugs or even DNA in gene therapy, so that they can be released in a controlled manner through the membrane \citep{RabanelEtAl09, BarthesBiesel2016}. Applications encompass food, cleaning, cosmetic and pharmaceutical products \citep{ColeEtAl08}. For example, targeted drug delivery relies on the release of a capsule's content at a specific destination, requiring capsules suspended in liquid to travel through complex, branched vessel geometries \citep{Abkarian2008}. Biological cells are examples of naturally occurring capsules. Haemodynamics in the capillary network \citep{Popel2005, Balogh2018} and porous media such as the placenta \citep{Igor2019} involve the transport and deformation of discrete red blood cells, with diagnostic and treatment of micro-circulation conditions relying on advances in the understanding of the resulting rheology. There is also considerable diagnostic potential in isolating circulating cancer cells, which are typically of similar size but more deformable than healthy white blood cells \citep{Geislinger2013, Lim2014}, or malaria-infected red blood cells, which stiffen by a factor of five due to the parasite that enters them \citep{Bow2011}. The transport of a capsule through a vessel bifurcation is a fundamental feature of all these applications. In this type of geometry, the extensional nature of the flow promotes large deformations of the capsule \citep{Loubens2015}, thus offering prospects for quantifying capsule deformability. Moreover, fluid-structure interaction influences capsule trajectories in the vessel bifurcation so that these complex geometries may be exploited to separate capsules according to their stiffness. 

 The flow-based transport of capsules in confined geometries has been extensively characterised in straight channels and capillaries both experimentally and numerically (e.g., \cite{BarthesBiesel2016, Pozrikidis2005, Kuriakose2013}). Most studies have focused on the limit of negligible inertia where capsule deformation results from the balance between viscous shear forces and elastic restoring forces, quantified by the capillary number $Ca$. In confined channels, spherical capsules elongate along the direction of flow for high values of $Ca$  and their rear buckles inwards to form a characteristic parachute shape \citep{Risso2006, Hu2012}. Ellipsoidal capsules may exhibit motions such as tank-treading \citep{Walter2011, Loubens2016}. If a deformable particle is set into motion from an off-centred position in the channel, it will tend to migrate towards the centre of the channel cross-section because of the interplay between particle deformation and the shear profile. A wall-lift force can be generated due to an upstream-downstream flow asymmetry resulting from the wall-induced deformation of the particle, but migration also arises in unbounded Poiseuille flow as demonstrated for capsules by \cite{Doddi2008} and for vesicles (droplets encapsulated by an inextensible lipid bilayer membrane) by \cite{CoupierEtAl08, Kaoui2008}.  Flow in straight tubes or square channels is routinely employed to determine the shear modulus of thin-walled capsules through comparison of their deformation with numerical models \citep{Hu2013, Chu2013, Lefebvre2008}. Extensional flows that feature a stagnation point have also been used to measure the constitutive behaviour of capsules by trapping and deforming them in device geometries such as four-roll-mills, cross-junctions and T-junctions \citep{Lee2007, Loubens2014}.
 
A recent surge in computational effort has focused on characterising the flow of capsules through these more complex geometries, mostly three-dimensional. The capsule is typically modelled as a liquid droplet enclosed by a thin, hyperelastic membrane. \cite{Zhu2015} considered the flow past a right-angled corner, while \cite{Koolivand2017} examined the deformation of a capsule trapped at the stagnation point of a T-junction, with both geometries offering new measurement methods of the elastic properties of the membrane based on large deformations of the capsule. Following the two-dimensional numerical model of \cite{Wolfenden2011}, the motion of capsules through branched tubes has recently been addressed in three-dimensions by \cite{Wang2016, Wang2018}, who focused on the path selection of the capsule in the presence of inertia with a view to design an enrichment device. A similar geometry has also been considered by \cite{Villone2017} to design a deformability-based sorting device using numerical modelling. Although experiments have been conducted on blood flow in branched capillaries, e.g., see \cite{Popel2005, Pries1996}, there is a distinct lack of controlled single-capsule experiments due the challenge of reliably manufacturing objects of controlled shape, with elastic properties which can be characterised independently of the flow experiment. 

In this paper, we focus on millimetric ovalbumin-alginate capsules and soft beads, and we explore their dynamics as they travel down a channel that splits into two identical daughter channels at a T-junction. We find that for our capsules (which have wall thicknesses up to 26\% of the radius and variable pre-inflation), the stiffness is best quantified by the force required to deform the capsule e.g., by compressing it between parallel plates to 50\% of its initial diameter \citep{CarinEtAl03, Risso2004, RachnikEtAl06}. We show that the capsule dynamics in the flow depend solely on a capillary number based on this force, which captures the non-linear elastic properties of the encapsulating membrane. We do not attempt to trap capsules at the stagnation point of the T-junction, and focus instead on their transport through the device. We find that their maximum extension in the T-junction offers a simple and sensitive measure of their relative stiffness. The T-junction is advantageous over straight conduits in that the presence of an extensional flow results in larger deformations of the capsules. In contrast to static compression or trapping in a four-mill device, a T-junction does not require careful time-consuming micro-manipulation so that multiple capsules can be characterised in rapid succession. 

Techniques for the flow-based separation of cells and capsules according to stiffness are currently emerging, and are often based on methods employed for the size-based separation of rigid, spherical particles which have been extensively characterised \citep{Sajeesh2014}. Amongst these, passive methods rely primarily on the device geometry and the hydrodynamic interactions between particles and the device, and include deterministic lateral displacement devices (DLD) consisting of arrays of pillars \citep{HuangEtAl04, LongEtAl08}, pinched flow fractionation (PFF) where a particle-laden flow is ``pinched''  against a rigid boundary by a jet orthogonal to the main stream \citep{YamadaEtAl04, CupelliEtAl13}, and inertial separation techniques \citep{DiCarlo09}. Inertial separation techniques have also been applied to deformable particles (e.g., \cite{HurEtAl11}), albeit with lower efficiency. In the limit of low inertia, \cite{Biros2019} recently demonstrated separation according to stiffness of fixed-size vesicles in a DLD device with a two-dimensional numerical model. \cite{Zhu2014} used three-dimensional boundary integral simulations to show distinct paths taken by capsules of different elasticity when circumventing a single pillar. Their results have been confirmed and extended by recent experimental studies \citep{VesperiniEtAl17, Haener2017}. To our knowledge, sorting according to stiffness in branched geometries is limited to the numerical studies by \cite{Wang2016,Wang2018} and \cite{Villone2017} cited above. In the current paper, we explore the sorting capabilities of the T-junction for capsules of fixed size. We find that at the same driving flow rate, capsules of different stiffness follow distinct paths out of the T-junction and that the position from which the capsule is entrained into the daughter channel depends solely on $Ca$. By adding a diffuser that enhances initial capsule separation, we demonstrate that a T-junction can be used to sort capsules according to stiffness. In contrast, we find that a PFF device fails to separate fixed-size capsules according to their stiffness. 

The paper is structured as follows. The manufacture of capsules, their characterisation and the flow devices are presented in \S 2. In \S 3, we characterise the transport of capsules through the T-junction as they travel from the parent channel into one of the two daughter channels.  In \S 4, we build on the findings in \S 3 to design a T-junction sorting device to separate capsules according to their stiffness. Conclusions are given in \S 5. 

\section{Experimental Methods}
\subsection{Ovalbumin-alginate capsules}
\subsubsection{Manufacture}
The capsules consisted of a liquid core encapsulated in a cross-linked ovalbumin-alginate membrane. They were manufactured using the method introduced by \cite{Levy96} (see Appendix \ref{appA} for details). In short, a solution in water of sodium-alginate, propylene glycol alginate (PGA) and ovalbumin was prepared, which was then added dropwise to a solution of calcium chloride in order to gel the drops into spherical beads. The beads were immersed in an alkaline solution to catalyse a transacylation reaction on the surface of the beads between the ester groups of the PGA and the unionised amino groups of the ovalbumin. This resulted in the creation of amide bonds and the formation of a semi-permeable cross-linked membrane encapsulating the beads of diameter $D_{\rm bead}$. Following neutralisation of the alkaline solution, the gel core of the beads was dissolved by immersing them in a sodium citrate solution. The newly manufactured capsules were stored in a saline solution (11 g/l NaCl) and reached equilibrium after approximately 24 hours. During this period, water permeated through their membrane, resulting in the inflation of the capsules to a size larger than the initial bead radius by up to 30\%, depending on membrane elasticity and initial size, implying the presence of a significant pre-stress in the capsule membrane. Solid beads were made in a similar manner to capsules, except that their core was not liquefied by exposure to sodium citrate.

Following the production of a batch of capsules, the capsules were first sorted by visual inspection and those without defects were coloured with a 10g/l Rhodamine B solution ($\ge 95 \% $, Sigma-Aldrich). Compression tests (see \S \ref{sec:character}) before and after capsule dyeing showed that the dyeing had no measurable effect on capsule deformability. The capsule diameter $D$ and thus its inflation $D/D_{\rm bead}$, its sphericity (ratio of the minimum to maximum diameter $D_{\rm min}/D_{\rm max}$) and its membrane thickness $h$ were measured from top and side-view images. Capsules with the desired size and sphericity ($>90\%$) were selected for compression tests followed by flow experiments.
The geometric properties of the capsules used in the present study are listed in Table \ref{table:Capsules}.

\begin{table}
	\centering
	\begin{tabular}{c|cccccc|c|cc}
		Capsule & Diameter & \multicolumn{2}{c}{Diameter } & Sphericity  & Membrane & Inflation & $F_{50 \%}$ \\
		 & (mm) & \multicolumn{2}{c}{Range (mm)}  & & thickness & $D/D_{\rm bead}$&& &   \\
		&$D$ &  $D_{\rm min}$ & $D_{\rm max}$  &$\displaystyle{\frac{D_{\rm min}}{D_{\rm max}}}$  & $2h/D$  & $\pm 5\%$     &  (mN)    \\
		&&&&& $(\pm 4\% )$&& \\ \hline
		TJ1 & 3.77   & 3.55 & 3.87  & 0.92& 0.18 & 1.3    & 2.4  \\
		TJ2 & 3.77  & 3.65 & 3.85 &  0.95 &  0.26 & 1.19   & 4.9  \\
		TJ3 & 3.87   & 3.69 & 3.94 & 0.94 & 0.16 & 1.20 & 9.2  \\
		\hline
		P1 & 3.89 & 3.71 & 4.04 & 0.91 & 0.12 & 1.20  & 1.9   \\
		P2 & 3.91 & 3.79 & 4.08  & 0.93 & 0.18 & 1.23  & 3.0    \\
		P3 & 3.93   & 3.76 & 4.08 & 0.92 & 0.19 & 1.15 & 8.4    \\
		P4 & 3.91   & 3.73& 4.03 & 0.93 & 0.18 & 1.23 & 17.1   \\
		\hline
		Elastic bead
		 & 3.90  & 3.77 & 4.22 & 0.89 & - & - &  33   \\ 
		 TJ4 &&&&&&& \\ \hline
	\end{tabular}
	\caption{Properties of the capsules and solid beads used in the experiments. The capsule diameter $D$, sphericity $D_{\rm min}/D_{\rm max}$, membrane thickness $h$ and inflation $D/D_{\rm bead}$, where $D_{\rm bead}$ is the diameter of the solid bead prior to liquefaction of its core, were measured from top and side-view images of the capsules. Capsules were compressed between parallel plates and the force required to compress them to 50\% of their initial height ($F_{50\%}$) was measured. 
	}
	\label{table:Capsules}
\end{table}

\subsubsection{Capsule characterisation} 
\label{sec:character}

\begin{figure}
	\centering
	\includegraphics[width = 1.0 \textwidth]{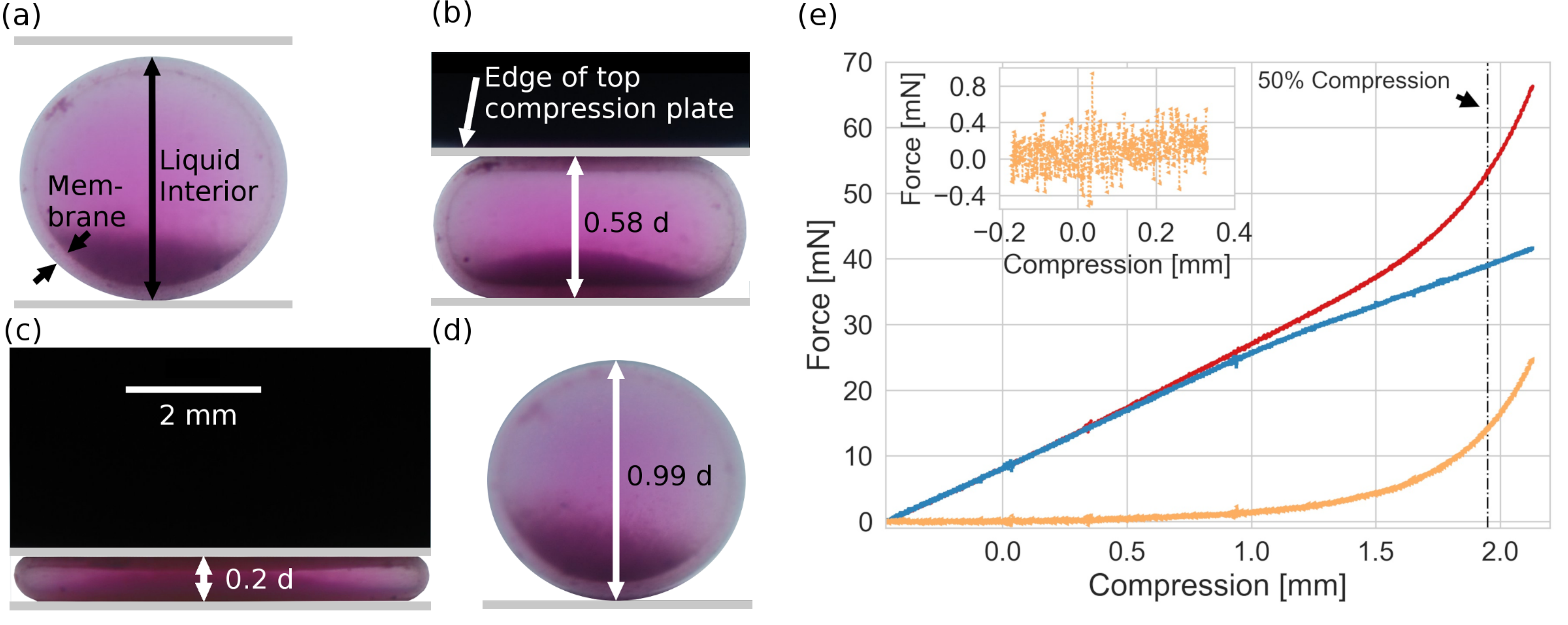}
	\caption{(a--d) Typical side-view images of a capsule immersed in a saline solution at four stages during the compression testing.
		(a) Before compression.
		(b) During compression.
		(c) Compression to 20\% of the capsule's initial height.
		(d) One second after the top plate has been returned to its initial position (see (a)). Note that the capsule has regained its undeformed shape with a height of 99\% of its initial value. 
		(e) An example of force versus compression data for capsule P4. The resistance to compression measured in the saline solution in the absence of a capsule (blue symbols) was subtracted from the force measurements in the presence of the capsule (red symbols) to yield the constitutive curve (orange symbols). The data shown is the average of three experiments. The inset shows a close-up of the force versus compression in the vicinity of the first contact with the compression plate. The point of contact is determined from images of the compression. The value of compression corresponding to 50\% of the initial capsule diameter is indicated with a vertical dash-dotted line.}\label{fig:egCompression}
\end{figure}

The constitutive relation that governs capsule deformation was measured by compression testing between parallel plates. An Instron 3345 Single Column Testing System (5~N load cell, accuracy $ \pm 0.5 $ mN) was used to measure the force exerted by a top plate lowered quasi-statically to compress a capsule placed on an anvil within a saline bath. Figures \ref{fig:egCompression}(a-d) show images taken during a typical compression experiment. Initially, the capsule is undeformed (a). The top compression plate is lowered, increasingly compressing the capsule (b,c) and then retracted (d). For the capsules used in the experiments, the maximum compression was 50\% of its initial height but in the example shown in Figure \ref{fig:egCompression}(c), the capsule is compressed by 80\%. Figure \ref{fig:egCompression}(d) shows that the capsule has returned to 99\% of its undeformed height following the return of the top plate to its initial position in (a), and thus demonstrates that these capsules can undergo large deformations elastically.

Figure \ref{fig:egCompression}(e) shows an example of force versus compression data recorded for capsule P4 (orange symbols). The top plate was lowered at a rate of 0.01 mm/s so that the compression test took approximately 5 minutes. This rate was sufficiently slow for the deformation to be quasi-static while fast enough to avoid osmotic changes. Doubling and halving the compression rate did not affect the measurement, confirming the rate independence of the process. The first point of contact between capsule and top plate was determined with a precision of $\pm 0.01$ mm by interpolating between side-view images taken every $2$~s. The resistance to compression measured in the saline solution in the absence of a capsule was subtracted from the force measurements to yield the constitutive curves (force versus upper plate displacement) of the capsules. A numerical model of the compression of an inflated, thick-walled spherical shell, developed using the open-source finite-element library {\tt oomph-lib} \citep{oomph-lib} in order to relate the experimental constitutive curves to the non-linear constitutive relation of the capsule membrane, suggests that the best fit across all capsules was provided by a Yeoh model \citep{Haener2017}.

The large deformations routinely observed when capsules were propagated in a flow meant that the non-linear form of the constitutive law was necessary to characterise their deformation. In addition, the capsules feature not only differences in the elastic modulus of the encapsulating membrane, but also in pre-inflation and in the thickness of this membrane (with values up to 26\% of the capsule radius, so that a membrane model is not appropriate). Hence, we quantify the resistance of a capsule to deformation by the force needed to compress it to 50\% of its initial height, $F_{50\%}$  (see Table \ref{table:Capsules}). This value of compression is indicated in figure \ref{fig:egCompression}(e) with a vertical dash-dotted line.

\subsection{Flow experiments}
\label{sec:EM}
\subsubsection{Experimental set-up}
\begin{figure}
	\centering
	\includegraphics[width = 1.0 \textwidth]{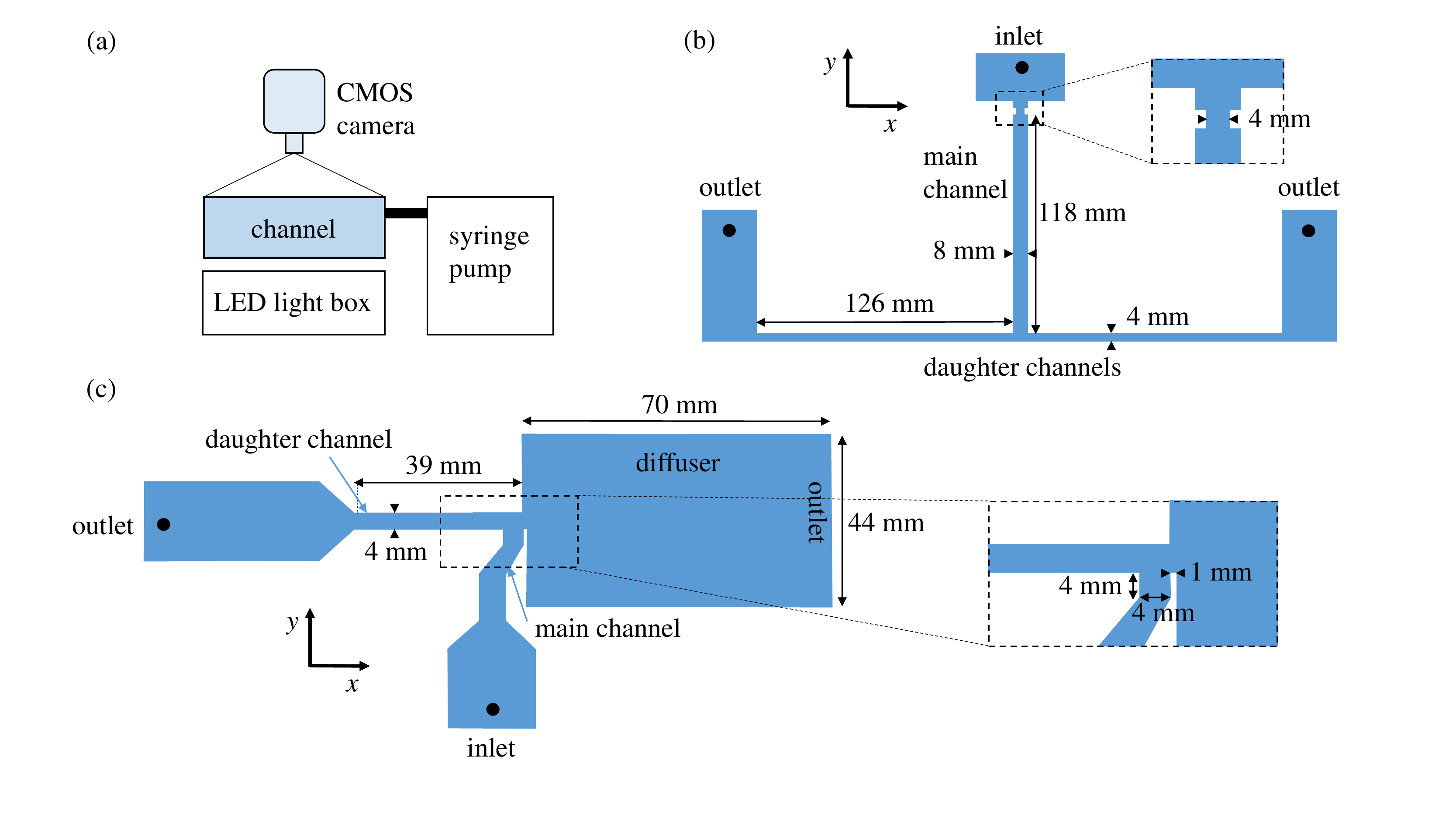}
	\caption{a) Schematic side-view diagram of the experimental setup; b) T-junction flow device; c) T-junction sorting device.}\label{fig:experiment}
\end{figure}

The experimental set-up used for the flow experiments is shown in Figure \ref{fig:experiment}(a). It consists of a flow channel levelled horizontally to within $0.5^{\circ}$ and back-lit with a custom-made LED light box. Capsule propagation was recorded in top-view at frame rates between 1 and 250  frames per second, using a monochrome CMOS camera (PCO, 1200hs) and a 50 mm micro-lens (Nikon) mounted vertically above the experiment.
The capsules were propagated in degassed silicone oil (polydimethylsiloxane, Dow Corning, 5000~cS) of viscosity $ \mu = 5.23 \pm 0.04$ kg~m$^{-1}$s$^{-1}$ and density $\rho= 970 \pm 10$ ~kg~m$^{-3}$ measured at the laboratory temperature of $21^{\circ} \pm 0.5^{\circ}$C. The capsules were approximately neutrally buoyant, with a downward vertical drift velocity of less than 0.2~mm~min$^{-1}$ when freely suspended in a beaker of silicone oil. A constant flow rate was imposed by injecting silicone oil into the inlet of the flow device using a syringe pump (KDS410, KD Scientific) fitted with a 50~ml stainless steel syringe (WZ-74044-36, Cole-Parmer). A second syringe pump (KDS210a, KD scientific) with two 20~ml syringes (gastight series, Hamilton) was used in the sorting experiments where two fluid inlets were required. In all the flow experiments, the Reynolds number was $Re= \rho\, Q \,W /(\mu A) <  9 \times 10^{-2},$ where $W$ is the width and $A$ the cross-section of the main channel. 
The propagation of capsules in flow was characterised by an elastic capillary number measuring the ratio of viscous to elastic forces and defined as
\begin{equation*}
Ca \ = \mu \frac{ Q}{A} \frac{D}{ F_{50 \%} }.
\label{eq:Ca}
\end{equation*}


The flow channels (Figures \ref{fig:experiment}(b,c)) were machined out of cast acrylic sheets (Perspex, Gilbert Curry Industrial Plastics Co Ltd.), using a CNC milling machine. Each device consisted of two facing plates screwed together. Each plate was milled flat to within  25~$\mu$m prior to the milling of the channels. The top plate featured fluid inlets and reservoirs while the millimetric channels were milled into the bottom plate. 

The T-junction flow device (Figure \ref{fig:experiment}(b)) consisted of a main channel bifurcating into two ``daughter'' channels of equal cross-section and length, oriented at $90^{\circ}$ to the main channel. The main channel leading to the T-junction had a cross-section of $(7.81 \times  3.98) \pm 0.02$ mm$^2$ and the cross-sectional dimensions of the daughter channels were $(3.8 \times 3.98) \pm 0.02$ mm$^2$. 

The T-junction sorting device (Figure \ref{fig:experiment}(c)) consisted of a main channel that branched into two daughter channels oriented at $90^\circ$ with respect to the main channel. The main channel and the left daughter channel both had the same cross-sectional dimensions of  $A= (3.8 \times 3.98) \pm 0.02$ mm$^2$, whereas the right daughter channel had a 1~mm long setion of cross-section of $A= (3.8 \times 3.98) \pm 0.02$ mm$^2$ followed by a 70 mm long section of cross-section $(39.8 \times 3.98) \pm 0.02$ mm$^2$, which we refer to as the diffuser. We note that the width of the main channel was approximately equal to the diameter of the capsule and thus, a centring mechanism was not needed in this device.  



\subsubsection{Experimental procedure}

Before inserting a capsule into the liquid-filled flow channel, any water on its outer surface was removed with a paper towel. The difference between the refractive indices of water and silicone oil makes water easily identifiable in the experimental images. If a water film initially coated the capsule, it could be isolated and removed  from the experiment by pushing the capsule through the flow channel at a high flow rate. At the start of each experiment, a capsule was positioned at the inlet of the device. A constant flow was then imposed which propagated the capsules through the T-junction device. Up to 70 experiments were performed for each capsule at each value of the flow rate. Thus, unless otherwise specified, the measurements shown in \S \ref{results} are average values with error bars determined from the unbiased standard deviation of the dataset.

In the sorting device, the distance of the centroid of the particles from the centreline of the main channel was less than 1.6\% of the channel width because of the strong confinement of capsules in the channels leading to the T-junction. In contrast, in the flow device, capsules were aligned along the centreline of the main channel using a short stepwise constriction of the channel to a width of 4~mm near the inlet (see figure~\ref{fig:experiment}(b)). This resulted in a deviation of the centroid of the particles from the centreline of the channel of less than 3.8\%, with a mean of the deviation of all particles of $0.6\%$. The symmetry of the flow device was estimated with a sample of 356 experiments: $53 \%$ of capsules turned into the right daughter channel. Using the estimator of true probability $\text{EP}$, a standard deviation corresponds to $\text{EP} = (2 \sqrt{n})^{-1} = 3 \%$. Therefore, the junction is within a standard deviation of being symmetric. We note that the stiffer solid beads (TJ4) exhibited greater variability in centring compared with the softer capsules. This is because the enhanced deformation of softer particles is associated with larger non-inertial lift forces that displace them faster away from boundaries, and thus promote more effective centring within the channel \citep{Doddi2008}.
The flow in the T-junction device exhibits a stagnation point, so that a perfectly centred and spherical capsule under ideal flow conditions could get trapped at this stagnation point. However, in the experiments, all capsules propagated in finite time because the small imperfections in the imposed flow, device and capsule break the symmetry of the flow about the centreline of the main channel. 


\subsubsection{Data analysis}

For each experimental run, between 200 and 1000 images were recorded with a resolution of $800\times 500$ pixels. We used Python 2.7 and OpenCV ({\tt https://opencv.org/}), a cross-platform, open-source computer vision library, for the image analysis. We applied background subtraction with either an adaptive threshold or a simple uniform threshold determined via the Otsu method \citep{Otsu79}. In the resulting black and white image, all contours were identified with a Canny filter \citep{Canny86}.

The visualisation in top-view yields images of the capsules in an $x$--$y$ plane of view averaged over the depth of field (see Figure~\ref{fig:experiment}(b) for coordinate system). Perimeter, area and centroid of the projected area of the capsules were determined from the contours extracted from these images. In the experiments reported here, the perimeter usually consists of 400 pixels and the centroid accuracy is better than 0.1 pixels. The accuracy was determined by creating test images of a perfect sphere with a given resolution and analysing the resulting images. The linear dimensions of the capsules along the $x$ and $y$ directions were measured by fitting an enclosing box to the outline.

\section{Capsule motion and deformation through the T-junction flow device}
\label{results}

  \subsection{Capsule motion through the T-junction}
 \label{sec:TJ-Traj}
 \begin{figure}
 	\centering
 	\includegraphics[width = \textwidth]{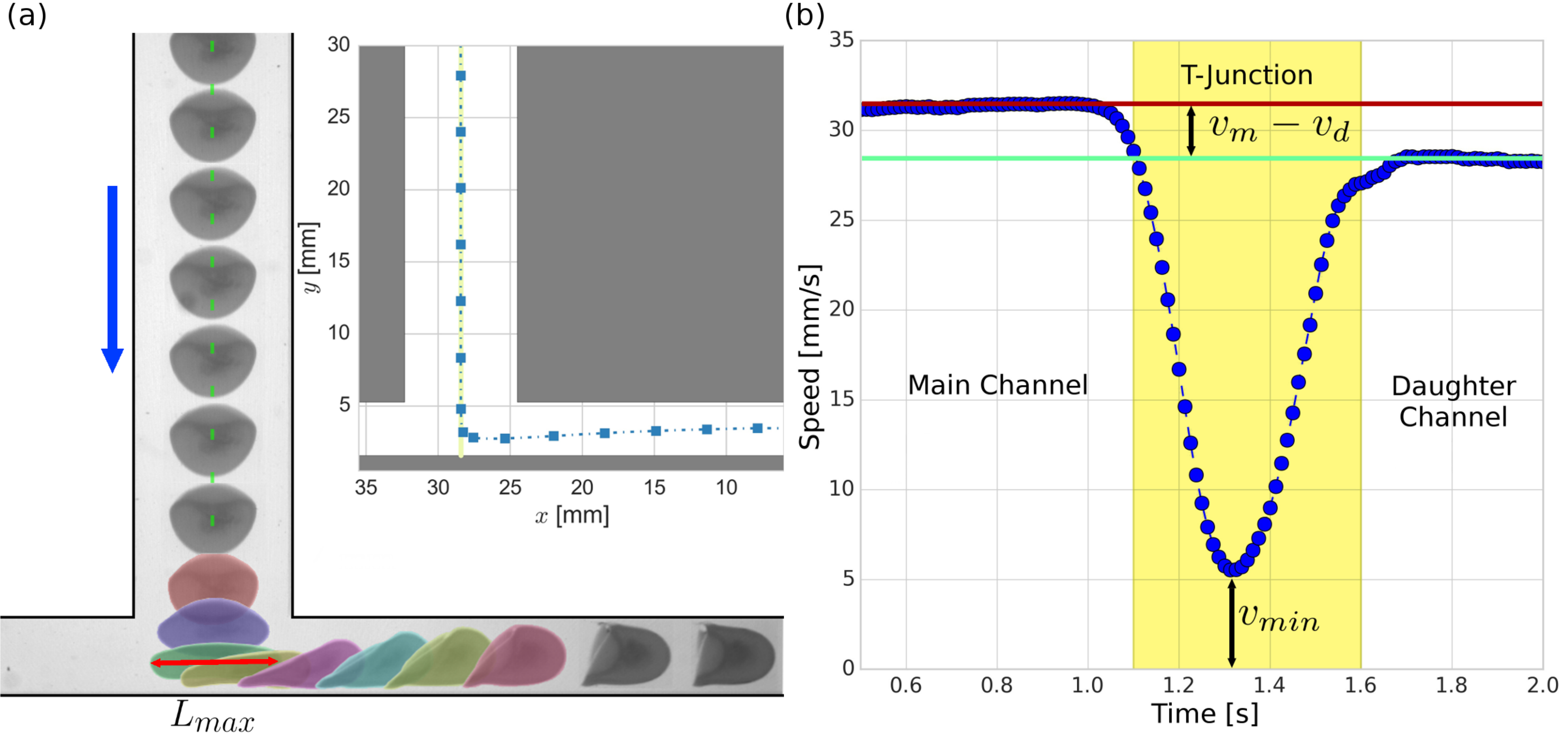}
 	\caption{(a) Capsule TJ1 travelling through the T-junction at $Ca = \left(1.7 \pm 0.4 \right) \times 10^{-1}$ ($ Q = 40$ {\normalfont ml/min}). Subsequent images have been overlaid to illustrate the dynamics; selected frames were coloured to enhance visibility. The centreline, about which the setup is symmetric, is shown with a green dashed line. The inset shows the trajectory of the centroid of the projected area of the capsule in the $x-y$ plane as a blue dotted line with every 10th point shown as a blue square. The average $x$ position of the capsule centroid measured between 4 and 10 capsule radii before the start of the T-junction is shown as a yellow line.
 		(b) Smoothed speed of capsule TJ3 passing through the T-junction at a flow rate of
 		$5$~{\normalfont ml/min}, $Ca = \left(0.56 \pm 0.03 \right) \times 10^{-2}$. } \label{fig:egImg}
 \end{figure}
 
 A typical example of the motion of a capsule (TJ1) through the T-junction is shown in Figure \ref{fig:egImg} in terms of its position (a) and speed (b) for $Ca = 0.17 \pm 0.04$. The trajectory of the capsule centroid is shown in the inset of Figure \ref{fig:egImg}(a) as a dotted blue line with every tenth measurement point marked by a blue square. The centroid position upon exit of the centring device (at the upstream end of the main channel, position indicated by a yellow line) is shifted by less than $0.02$ mm (0.5 \% of the channel half-width) from the centreline of the channel (dashed green line) and thus, the capsule follows a straight path through the main channel into the junction. 
 The speed of the capsule in Figure~\ref{fig:egImg}(b) was determined by finite-differencing its centroid position with respect to time and was smoothed with a weighted moving-average filter. The speed in the main channel, $v_m$, is approximately constant but in the T-junction, the capsule decelerates while it travels along the $y$-axis to reach a minimum speed, $v_{min}$, close to the bottom wall of the junction. In 15\% of the experiments, the capsules got trapped at the stagnation point in the T-junction for prolonged periods during which $v_{min} \simeq 0$~mm/s. As the capsule approaches the bottom wall, it extends in the $x$ direction to reach a maximum length $L_{max}$ (indicated with a red line segment in the figure) and compresses in the $y$-direction.  Soft capsules extend considerably for sufficiently large flow rate (Figure \ref{fig:egImg}(a)) and develop a dumbbell shape reminiscent of the early extension of bubbles prior to break-up \citep{FuEtAl11, Dawson2015}. The deformation of the capsule means that its centroid can approach the bottom wall of the junction more closely than the centroid of a rigid particle which retains its spherical shape. 
 
After reaching its minimum speed, the capsule accelerates into the daughter channel to reach a constant speed $v_d$, as shown in Figure \ref{fig:egImg}(b). 
The value of  $v_d$ is lower than the speed in the main channel, $v_m$, because of the enhanced confinement in the daughter channel (which has a width of $4.0$~mm, similar to the diameter of the capsule and half of the width of the main channel).  Once inside the daughter channel, the capsule migrates towards the centreline over distances of several capsule diameters due to viscous shear forces, and regains symmetry about the centreline of the channel towards the end of the visualisation window (Figure \ref{fig:egImg}(a)). 
 
 Figure \ref{fig:egImg}(a) shows that the rear of the capsule adopts a parachute shape in both main and daughter channels, consistent with previous studies \citep{Risso2006, Lefebvre2008, Hu2013}. We find that its rear is deformed into a concave shape beyond a threshold value of the capillary number of $Ca_c = 0.07 \pm 0.02$ in the main channel, while in the daughter channel, $Ca_c = 0.05 \pm 0.02 $. The increased viscous drag in the daughter channel translates into the increased deformation (more indented parachute shape) of the re-centred capsule compared to the capsule in the main channel. 


 \subsection{Capsule speed in the straight sections of channels}
 \begin{figure}
 	\centering
 	\includegraphics[width = \textwidth]{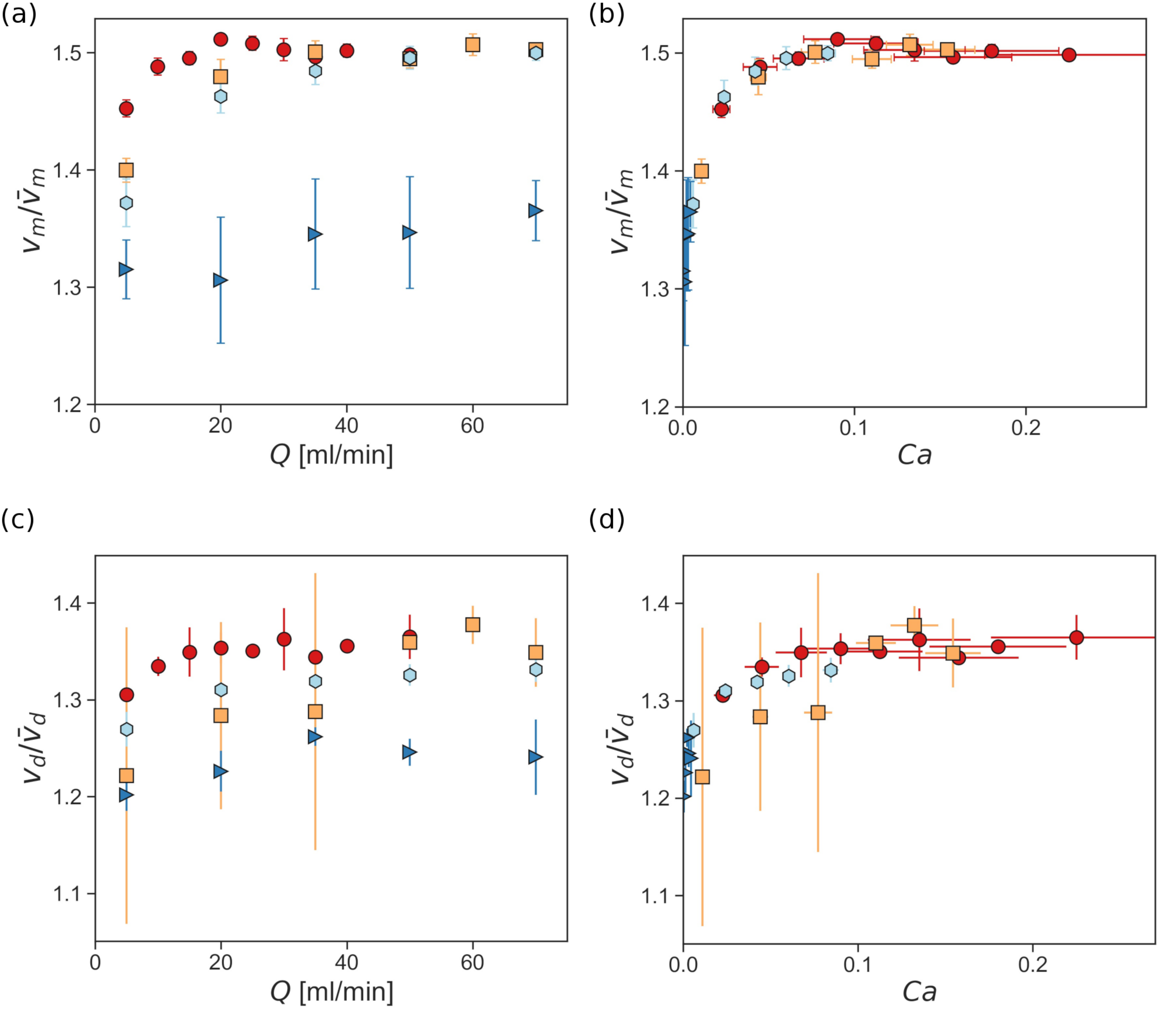}
 	\caption{Ratio of capsule speed to average flow velocity as a function of flow rate $Q$ (a, c);  capillary number $Ca$ (b, d). Results shown in (a,b) are for the main channel, and in (c,d) for the daughter channel. The symbols represent
 		capsules TJ1 (red circles), TJ2 (orange squares), TJ3 (light blue hexagons) and the solid bead TJ4 (blue triangles).  All the capsules and the solid bead are of similar size with $3.77$~mm $\le D \le 3.87$~mm. The data collapse onto a master curve when plotted against $Ca$, which confirms that $Ca$ governs the motion of capsules of fixed size in a straight channel.} \label{fig:Speed}
 \end{figure}
 
\begin{figure}
 	\centering
 	\includegraphics[width = \textwidth]{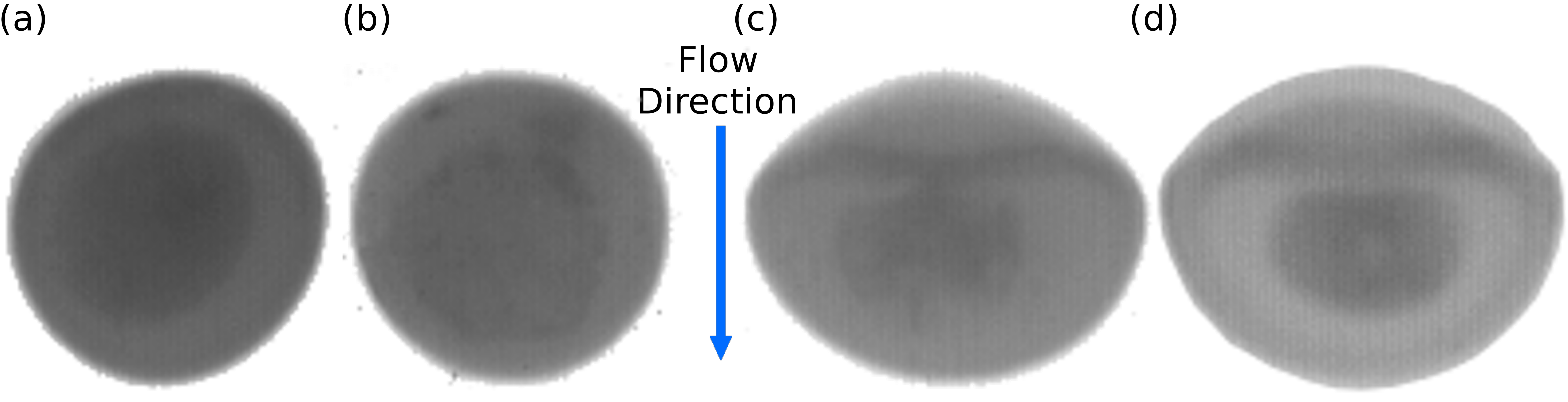}
 	\caption{Capsules of different stiffness show similar deformation at the same value of $Ca$ in the main channel. The arrow indicates the direction of flow. 
 		(a) TJ3, $Q = 2$ {\normalfont {\normalfont ml/min}},   $Ca = \left( 2.2 \pm  0.1 \right) \times 10^{-2}$.
 		(b) TJ1, $Q = 5$ {\normalfont {\normalfont ml/min}}, $Ca = \left( 2.1 \pm 0.5 \right) \times 10^{-2}$.
 		(c) TJ1, $Q = 35$ {\normalfont {\normalfont ml/min}}, $Ca = \left( 15 \pm 3 \right) \times 10^{-2}$.
 		(d) TJ2, $Q = 70$ {\normalfont {\normalfont ml/min}}, $Ca = \left( 14 \pm 1 \right) \times 10^{-2}$.} \label{fig:CapsuleMC3p8}
 \end{figure}

 The ratio of capsule speed to mean flow velocity (fractional speed) is shown in Figure \ref{fig:Speed} as a function of flow rate and $Ca$, respectively, for propagation in both main (Figure \ref{fig:Speed}(a,b)) and daughter channels (Figure \ref{fig:Speed}(c,d)). Rigid spheres travel at speeds of one to two times the mean velocity of the fluid, depending on their size (see, e.g., \citet{QuddusEtAl08}).  The fractional speed of capsules depends on their deformation, and thus on flow rate, even if their size is fixed. In Figure \ref{fig:Speed}(a), the propagation of the solid bead (TJ4) corresponds to that expected for a near-rigid object: it travels at an almost constant fraction of the mean flow velocity over the range of flow rates investigated. In contrast, the fractional capsule speed increases with flow rate towards a single constant value larger than that measured for the solid bead. At low values of the flow rate (e.g. $Q=5$~ml/min), the fractional speed increases as the stiffness of the capsule decreases. The plot of the same data against $Ca$ in Figure \ref{fig:Speed}(b) indicates that the data collapse onto a master curve and thus, that the capillary number governs the motion of capsules (of fixed size) in the straight sections of channel. This suggests that the deformation of different capsules at the same capillary number is the same. This is confirmed in Figure \ref{fig:CapsuleMC3p8}, which shows that the capsules remain approximately spherical at low $Ca$ (Figure \ref{fig:CapsuleMC3p8}(a,b)), while at higher $Ca$, the capsules exhibit the characteristic parachute shape (Figure \ref{fig:CapsuleMC3p8}(c,d)). 
 
 In the daughter channels, the increased confinement is associated with an increased sensitivity to the initial sphericity of the underformed capsule, which results in larger scatter of the data (Figure \ref{fig:Speed} (c)). Despite this scatter, the fractional capsule speed approximately collapses onto to a master curve as a function of $Ca$ and appears to reach a plateau at higher $Ca$ (Figure \ref{fig:Speed} (d)). 
 
 We find that at large $Ca$ the capsules in the main and daughter channels travel at approximately $v_m/\bar{v}_m = 1.5$ and $v_d/ \bar{v}_d =1.35$, where $\bar{v}_m$ and $\bar{v}_d$ denote the mean fluid velocities in the main and daughter channels, respectively.
 This latter value is consistent with the numerical result of \citet{Hu2013} in a square channel for a capsule with a ratio of diameter to channel width of 0.95 (0.94 -- 0.97 in our experiments; see Table \ref{table:Capsules}), and a thin membrane characterised by a Skalak constitutive law, where the fractional speed increased from $1.2$ to $1.3$ over a representative range of $Ca$,  based on the shear modulus of the capsules.  Our experimental results for a Yeoh constitutive law yield a similar range of $1.2 \le v_{d}/\bar{v}_d \le 1.35$. This suggests that capsule propagation in straight sections of tube is not strongly influenced by the exact constitutive relation of the capsule. 
 
\subsection{Capsule deformation in the T-junction}
\label{sec:T-junction}
\subsubsection{Maximum deformation}
\begin{figure}
	\begin{center}
		\includegraphics[width = \textwidth]{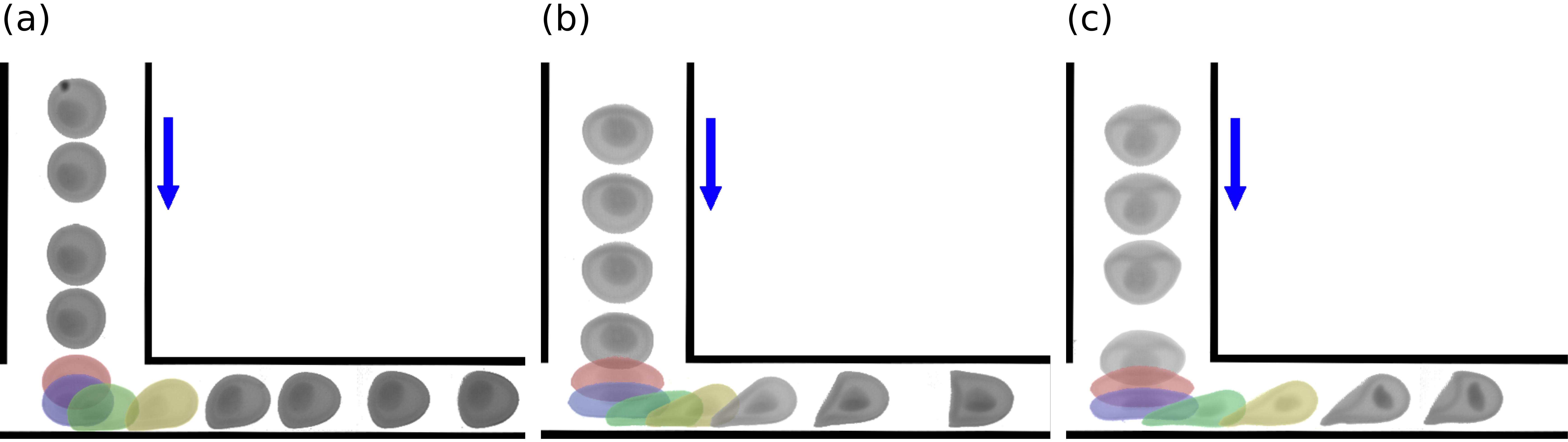}
	\end{center}	
	\caption{Series of snapshots illustrating the travel of capsule TJ2 through a T-junction for three different flow rates.
		(a) $Q=5$ {\normalfont {\normalfont ml/min}}, $Ca = \left( 1.1 \pm 0.1 \right) \times 10^{-2}$.
		(b) $Q=35$ {\normalfont {\normalfont ml/min}}, $Ca = \left( 7.7 \pm 0.8 \right) \times 10^{-2}$.
		(c)  $Q=70$ {\normalfont {\normalfont ml/min}}, $Ca = \left( 15.4 \pm 1.6 \right) \times 10^{-2}$.	}
	\label{fig:1Capsule}
\end{figure}

The deformation of a single capsule (TJ2) as it travels through the junction for three values of the flow rate is shown in Figure \ref{fig:1Capsule}. The overall deformation of the capsule in the T-junction increases with flow rate, leading to considerable shape changes particularly in Figure \ref{fig:1Capsule}(c). Similar deformation patterns were obtained for all the capsules tested. The membrane that encapsulates the liquid core routinely exhibits bending radii of the order of the (undeformed) membrane thickness, between 16\% and 26\% of the capsule radius, without suffering permanent deformation. 

\begin{figure}
	\begin{center}
		\includegraphics[width = \textwidth]{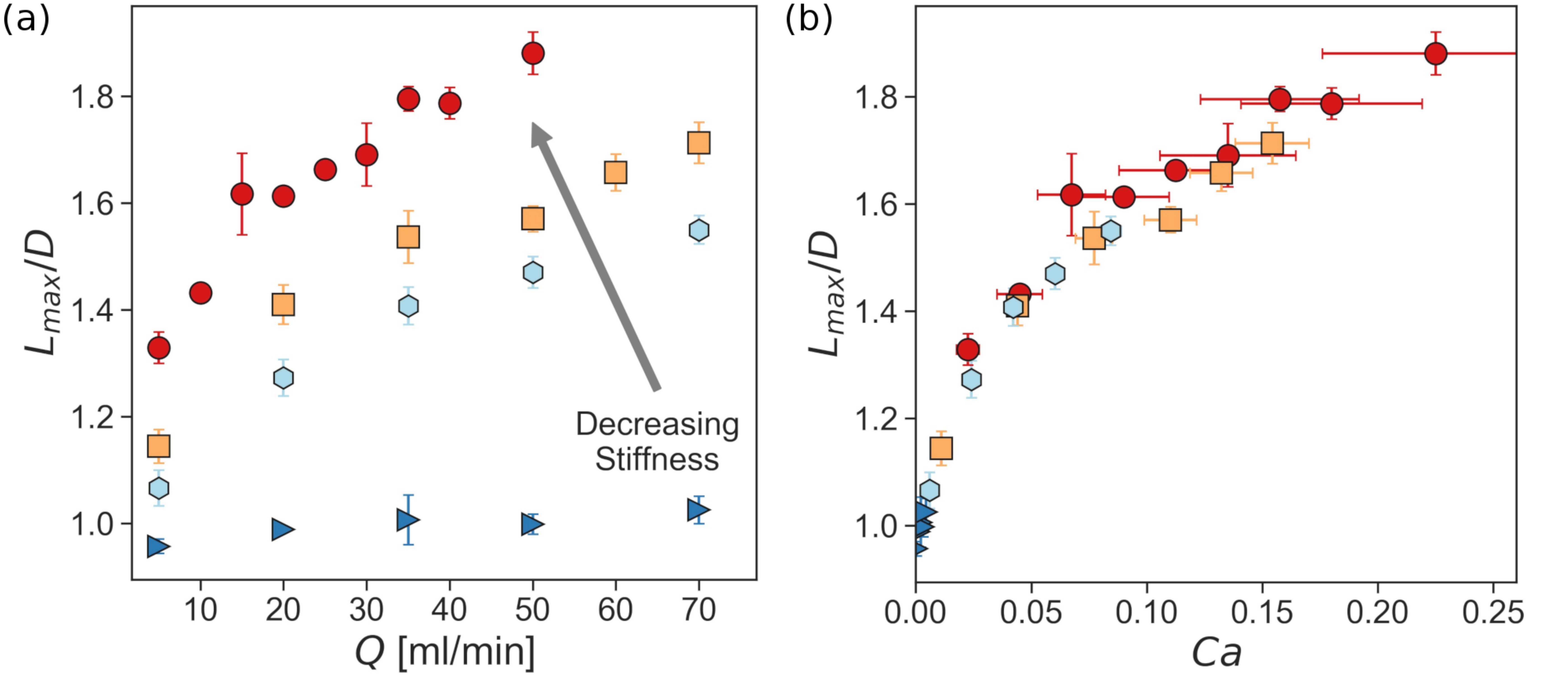}
	\end{center}
	\caption{Maximum length of the capsules in the T-junction as a function of flow rate $Q$ (a); $Ca$ (b). The symbols represent capsules TJ1 (red circles), TJ2 (orange squares), TJ3 (light blue hexagons) and the solid bead TJ4 (blue triangles). All the capsules and the solid bead are of similar size with $3.77$~mm $\le D \le 3.87$~mm. (b) indicates that $L_{max}$ depends uniquely on $Ca$ for fixed particle size to within experimental error. The symbols are the same as in previous graphs.} \label{fig:MaxExtension}
\end{figure}

The maximum length $L_{max}$ of three capsules and of the solid bead are plotted in Figure \ref{fig:MaxExtension}(a,b) as a function of $Q$ and $Ca$, respectively. Whereas the near-rigid solid bead barely deforms (<3\%) for the range of flow rates investigated, the capsules of different stiffness are clearly distinguishable based on their maximal length in Figure  \ref{fig:MaxExtension}(a). The most deformable capsule (TJ1) reaches the largest values of $L_{max}$ for all flow rates, while the stiffest capsule (TJ3) exhibits the least deformation. Figure \ref{fig:MaxExtension}(b) shows that the values of $L_{max}$ collapse onto a master curve to within error bars as a function of $Ca$. This indicates that $Ca$ uniquely determines the deformation of these capsules (which are of similar size) and the moderate error bars indicate that the method yields reproducible results. 

The significant variation of the maximum length as a function of $Ca$ makes it a promising metric for quantifying the relative stiffness of capsules in flow. At fixed flow rate, the three capsules ($2.4\; \text{mN}\, \le F_{50 \%} \le 9.2$~mN; see Table \ref{table:Capsules}) exhibit a measurable shift in their values of $L_{max}$. This indicates that factors of 2 and 4 in the force required to compress the capsule to 50\% of its initial height are easily resolved over the entire range of flow rates investigated. The data in Figure \ref{fig:MaxExtension} also suggest that the deformation of soft capsules is captured most accurately. At $Q=50$~ml/min; the softest capsule (TJ2)  has extended by $\sim 90\%$; TJ1 which is stiffer by a factor of two has extended by $\sim 50\%$; whereas TJ3 which is stiffer by a factor of four has extended by more than $40\%$. The resolution of this method for highly deformable capsules means that it complements the compression method commonly used to measure the stiffness of millimetric capsules \citep{CarinEtAl03, Risso2004, RachnikEtAl06}. 

\subsubsection{Change of direction within the T-junction}
\label{sec:turn}

\begin{figure}
	\begin{center}
		\includegraphics[width = \textwidth]{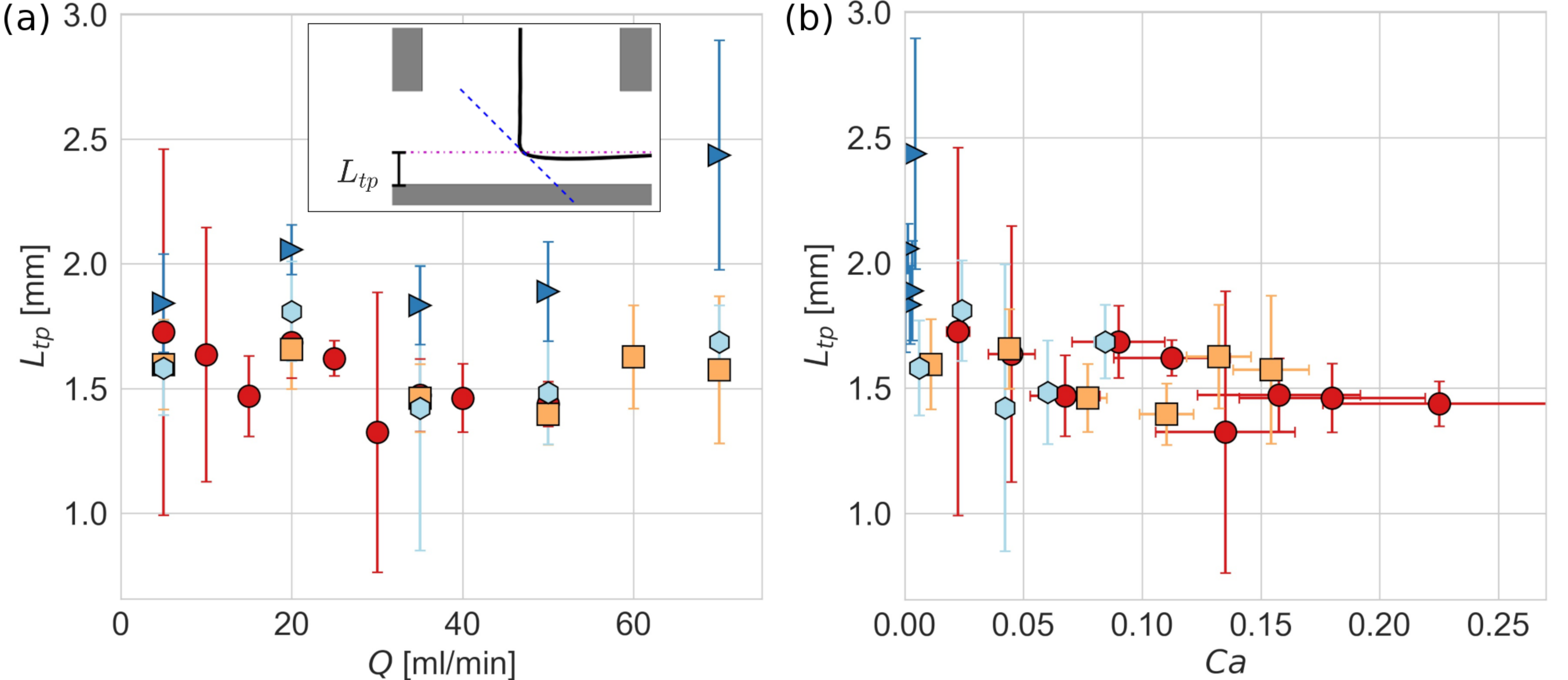}
	\end{center}
	\caption{
		Distance from the bottom wall of the T-junction at which the capsules turn into the daughter channel as a function of  (a) flow rate $Q$ and (b) capillary number $Ca$.  The symbols represent capsules TJ1 (red circles), TJ2 (orange squares), TJ3 (light blue hexagons) and the solid bead TJ4 (blue triangles). All the capsules and the solid bead are of similar size with $3.77$~mm $\le D \le 3.87$~mm. 
	} \label{fig:TpTJ}
\end{figure}

The change of direction of the capsule in the T-junction is a key feature of its trajectory. We will show in \S \ref{subSec:TJMode} that it enables the use of the T-junction as a device for sorting capsules according to their stiffness. To characterise the entrainement of the capsule into the daughter channel, we record the distance, $L_{tp}$, of the capsule centroid from the bottom wall of the T-junction when the velocity vector is angled at 45$^{\circ}$ to the centreline of both channels (see inset of figure \ref{fig:TpTJ}(a)). Figure \ref{fig:TpTJ} shows the dependence of $L_{tp}$ on $Q$ and $Ca$, respectively. The variation of $L_{tp}$ with $Q$ is small and the large error bars indicate significant variability between repetitions of the same experiment. However, the solid beads systematically turn at a greater distance from the bottom wall of the T-junction than the capsules. Figure \ref{fig:TpTJ}(b) shows a slight decrease of $L_{tp}$ with increasing $Ca$, with values of approximately 2~mm for the lowest values of $Ca$ while for $Ca = 0.2$, $L_{tp} < 1.5$~mm. This means that the more deformed capsules start their propagation into the daughter channel from a position closer to the bottom wall.



\subsection{Relaxation upon exit from the T-junction} \label{sec:Relaxation}

\begin{figure}
	\begin{center}
		\includegraphics[width = \textwidth]{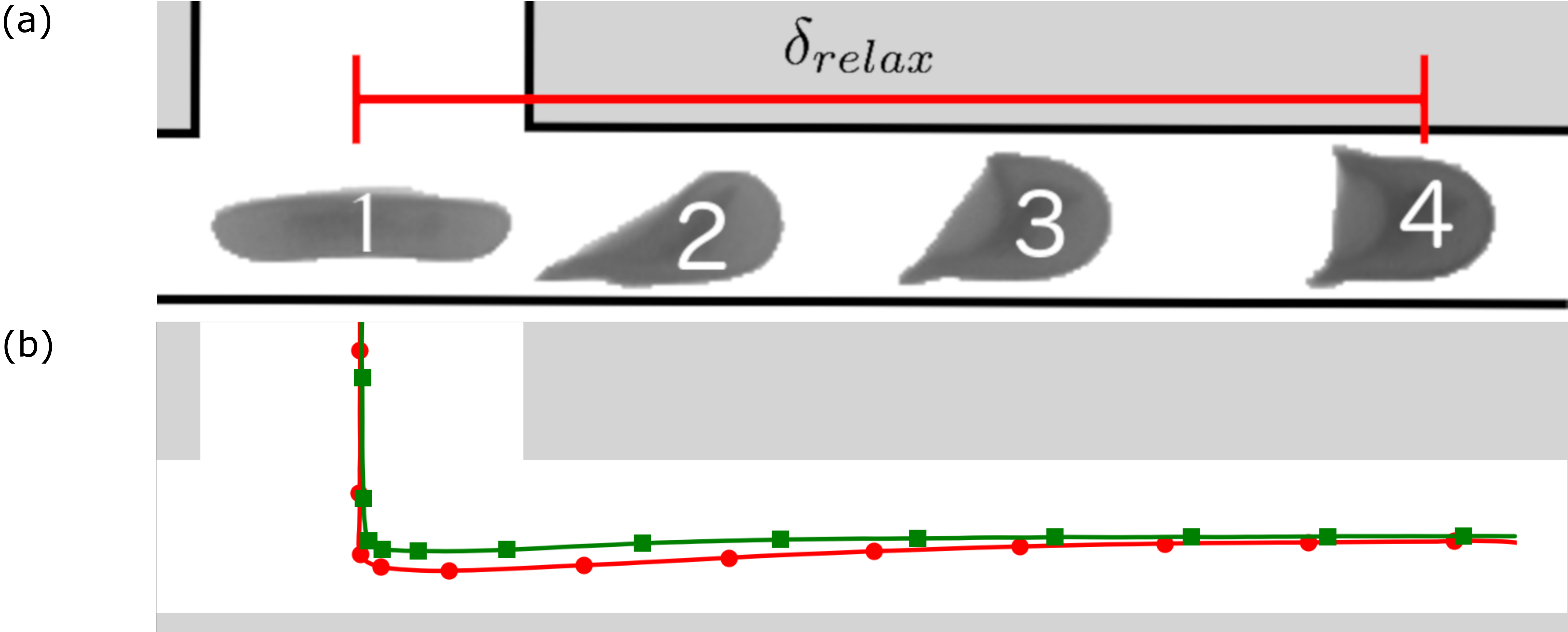}
	\end{center}
	\caption{
		(a) A series of images illustrating the evolution of capsule TJ1 in the daughter channel for $Q=50$~ml/min ($Ca = \left( 21 \pm 5 \right) \times 10^{-2}$). 
		(b) Trajectories of the centroids of capsule TJ1 (red circles) and solid bead TJ4 (green squares) as they propagate towards a steady state in the daughter channel.} 	 \label{fig:TrajRelax}
\end{figure}

Figure \ref{fig:TrajRelax}(a) illustrates the evolution of capsule TJ1 as it propagates along the daughter channel. When the capsule changes direction within the T-junction, it is highly deformed and its centroid is displaced from the centreline of the daughter channel towards the bottom wall. The elongated capsule (snapshot labelled~1 in Figure \ref{fig:TrajRelax}(a)) is subject to a shear gradient and to wall lift forces which incline its nose towards the centreline (snapshot~2) and force it to migrate towards the centreline \citep{Doddi2008, CoupierEtAl08}. Once the nose of the particle has reached the centreline (snapshot~3), elastic restoring forces promote relaxation of the capsule towards a symmetric shape (snapshot~4), but this process typically occurs over a distance of several capsule diameters. In contrast, the solid bead exhibits minimal deformation, turns into the daughter channel close to its centreline and reaches a steady mode of propagation after covering less than half the distance required by TJ1 (see Figure \ref{fig:TrajRelax}(b)). 


\begin{figure}
	\begin{center}
		\includegraphics[width = \textwidth]{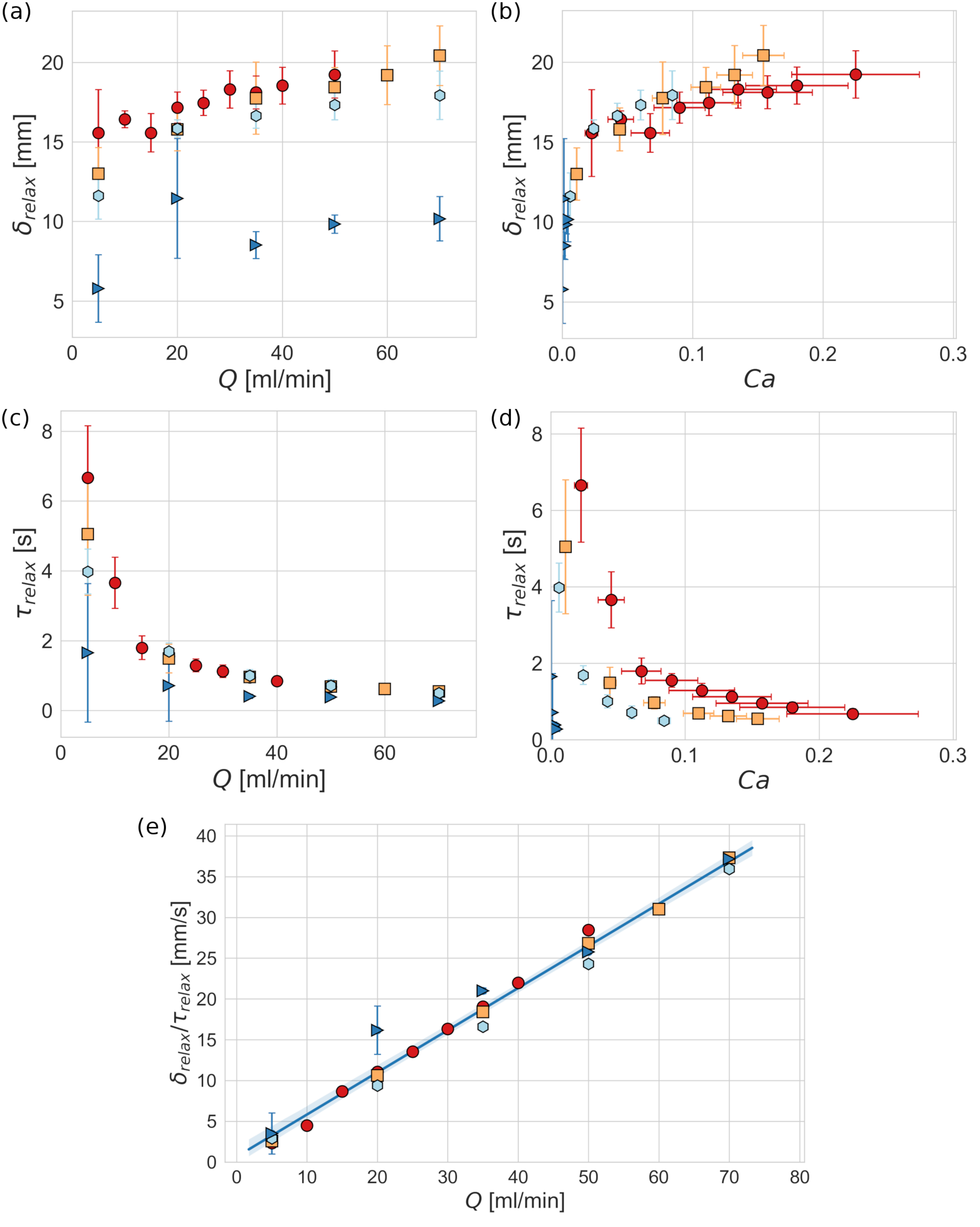}
	\end{center}
	\caption{ 
		Relaxation distance as a function of (a) flow rate $Q$ and (b) $Ca$. Relaxation time as a function of  (c) flow rate $Q$ and (d) $Ca$.
		(e) Average relaxation speed with the line of best fit shown in blue (the 95\% confidence interval is indicated by a light-blue shaded band around the line of best fit). The symbols represent capsules TJ1 (red circles), TJ2 (orange squares), TJ3 (light blue hexagons) and the solid bead TJ4 (blue triangles). All the capsules and the solid bead are of similar size with $3.77$~mm $\le D \le 3.87$~mm.}
	\label{fig:Relax}
\end{figure}

We define the relaxation distance $\delta_{\rm relax}$, and the associated relaxation time $\tau_{\rm relax}$, as the distance travelled by the centroid and time elapsed, respectively, from the point of maximum capsule deformation to the position at which the capsule reaches a state in the daughter channel where the transverse velocity of its centroid is $0.25$~mm/s (the limit of our resolution of the velocity). Results are shown as a function of flow rate in Figure \ref{fig:Relax}(a,c) for the three capsules and the solid bead previously shown in Figures \ref{fig:Speed} and \ref{fig:MaxExtension}. The relaxation distance increases with increasing flow rate and the relaxation time scale consistently decreases. Upon reduction of the stiffness, both $\delta_{\rm relax}$ and $\tau_{\rm relax}$ increase, which confirms that softer capsules travel further and longer before reaching a new steady state. When plotted as a function of $Ca$ in Figure \ref{fig:Relax}(b), $\delta_{\rm relax}$ collapses onto a master curve within error bars indicating that it is governed by the fluid-structure interaction in the system like the maximum length and the turning point plotted in Figures \ref{fig:MaxExtension} and \ref{fig:TpTJ}, respectively. In contrast, Figure \ref{fig:Relax}(c,d) shows that $\tau_{\rm relax}$ is not uniquely governed by $Ca$. This is because the time scale of capsule propagation is set by the imposed flow rate. In fact, Figure \ref{fig:Relax}(e) shows that the average speed of relaxation, $\bar{v}_{relax}= \delta_{relax}/\tau_{relax}$,  is directly proportional to the flow rate. This means that although the shape relaxation of softer particles is initiated from a more deformed state imposed by the flow pressure in the T-junction, they are also positioned closer to the wall of the daughter channel. Hence, the same flow in turn imposes larger wall lift forces, which return the capsules to the centreline with the same average speed. 


\newpage

\section{Sorting capsules according to their stiffness in a T-junction device}
\label{sorting}

\subsection{T-junction mode of operation}
\label{subSec:TJMode}

 \begin{figure}
	\centering
	\includegraphics[width = 0.75\textwidth]{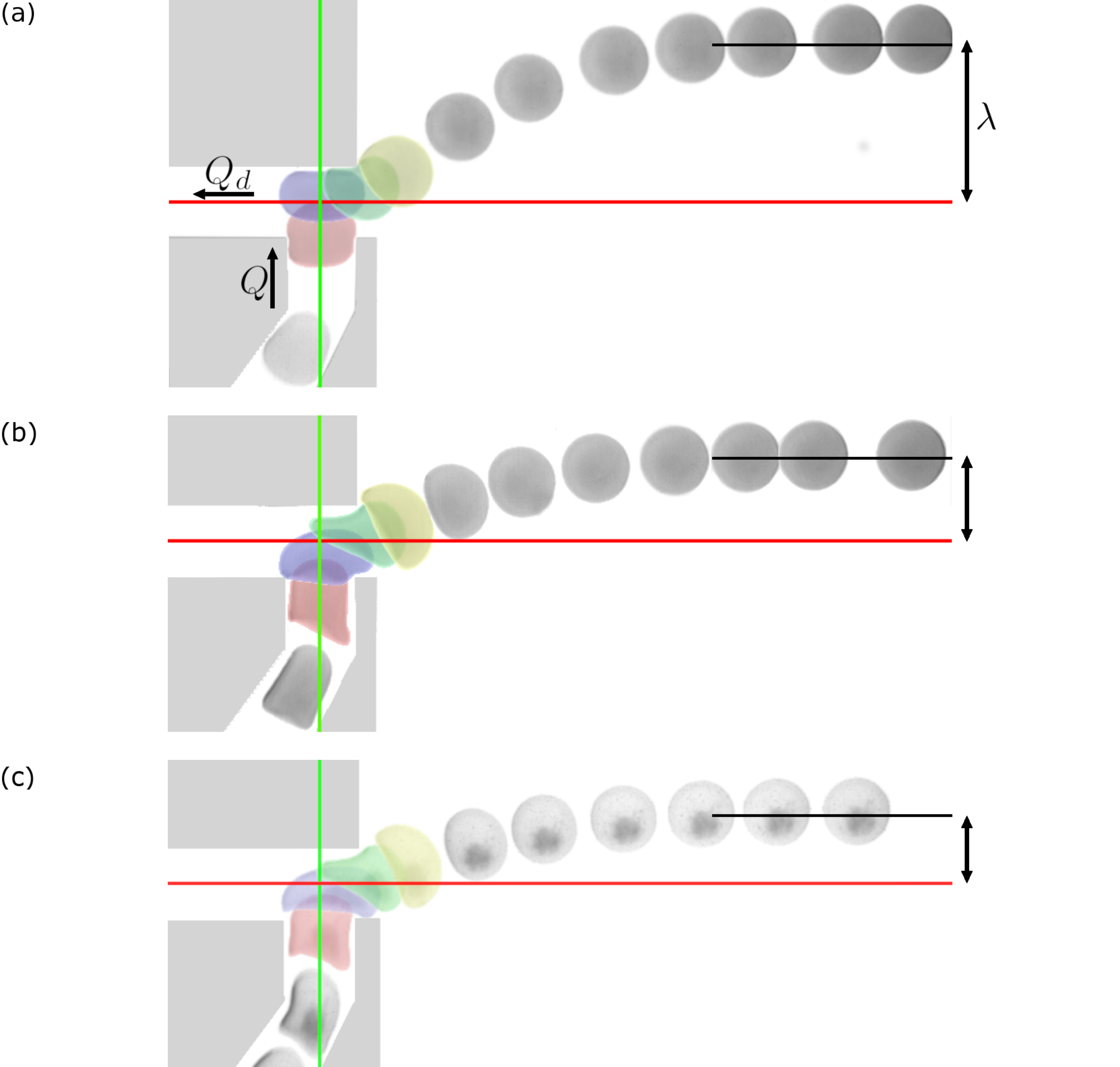}
	\caption{
		Sequences of images illustrating the propagation of capsules through the sorting device operated in T-junction mode. The red line indicates the centreline of the daughter channels and the green line that of the main channel. $\lambda$ is the offset of the capsule centroid from the centreline of the diffuser. The capillary number $Ca$ is based on the flow rate $Q$ and cross-section of the main channel. 
		(a) Capsule P4, $Q_{d} = 2.47$~ml/min, $Q =  5$~ml/min
		$Ca = (6.2 \pm 0.2) \cdot 10^{-3}$.
		(b) Capsule P4, $Q_{d} = 20$~ml/min, $Q =  40$~ml/min,
		$Ca =(5.0 \pm 0.2) \cdot 10^{-2}$.
		c) Capsule P1, $Q_{d} = 2.5$~ml/min, $Q =  5$~ml/min
		$Ca = (5.6 \pm 1.6) \cdot 10^{-2}$.
	}\label{fig:eg-PFF-TJ}
\end{figure}

\begin{figure}
	\centering
	\centering
	\includegraphics[width = 0.8 \textwidth]{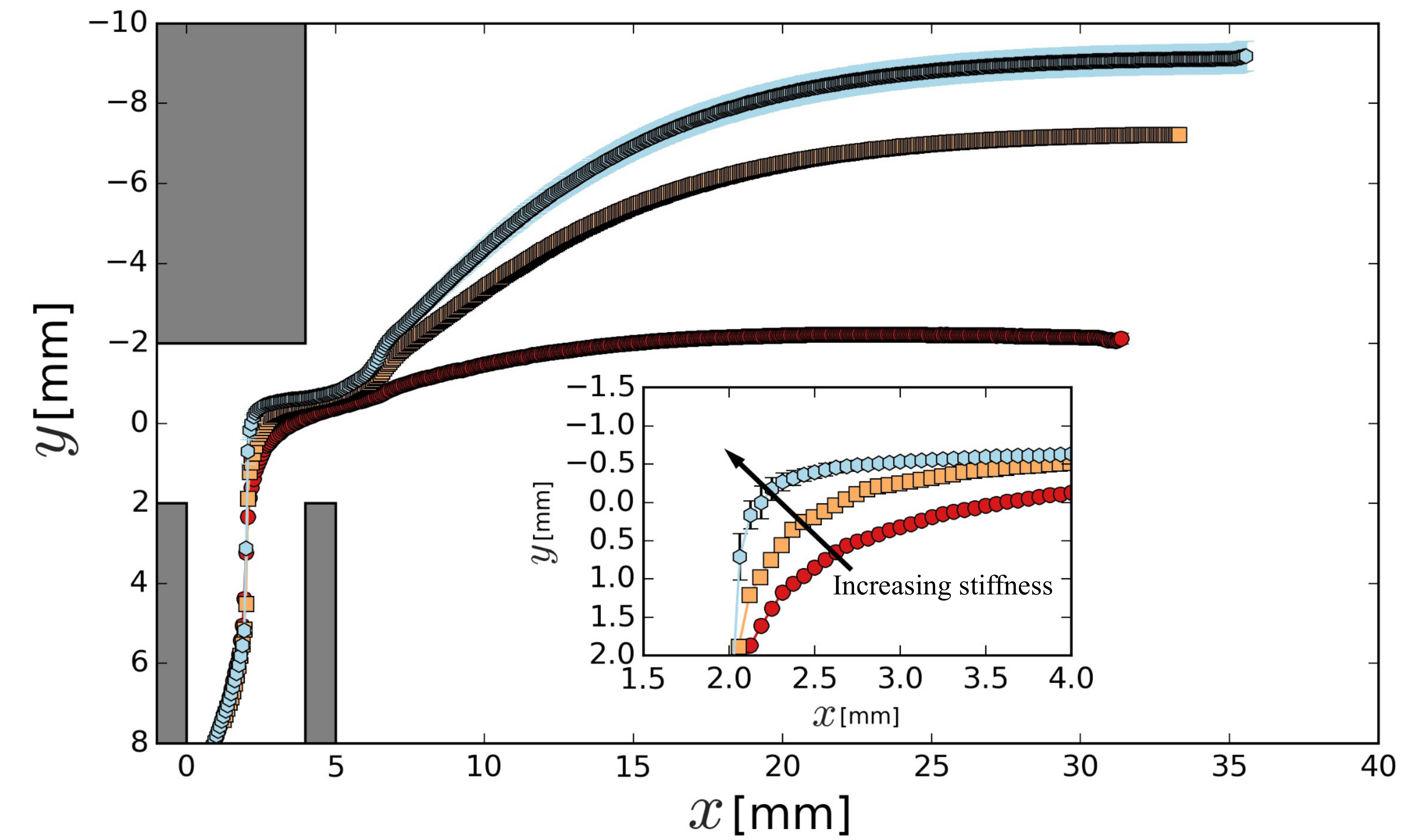}
	\caption{ Capsule trajectories and deformation in the sorting device operated in T-junction mode with a flow rate of $ Q = 10$~ml/min and $Q_d/Q=1/2$. Red circles: capsule P1; orange squares: P3; light blue pentagons: P4.  Average trajectories of the centroids of the capsules. The error bars indicate the standard deviation of multiple experiments. Inset: close-up view of trajectories in the T-junction. 
	}\label{fig:TrajAndWidth-PinTJ}
\end{figure}

The $Ca$-dependent deformation and separation of monodisperse capsules in the T-junction, demonstrated in \S \ref{sec:T-junction} and \S \ref{sec:Relaxation}, suggests that the T-junction may be exploited to passively sort capsules of fixed size according to their stiffness. For this purpose, the separation of capsules achieved within the T-junction needs to be enhanced in the daughter channels. In the geometry investigated so far, this separation is reduced post-junction by the convergence of the capsules onto the centreline of the daughter channel, which has a smaller cross-section than the main channel. We therefore turn to the investigation of a modified T-junction where one of the daughter channels is substantially enlarged to act as a particle diffuser in order to amplify capsule separation; see the schematic diagram of the T-junction sorting device presented in Figure \ref{fig:experiment}(c). Note that the main channel is straight only over the last 4~mm closest to the T-junction in order to reduce the thickness of the wall between the main channel and the diffuser to 1~mm, which is the smallest practically feasible value. Hence, the diffusing daughter channel (right hand-side of the T-junction) widens 1~mm downstream of its intersection with the main channel. Capsules were propagated through the main channel into the T-junction by imposing different values of the flow rate $Q$. A second flow rate $Q_d$ was imposed to control the withdrawal of fluid from the left daughter channel. All experiments were performed for $Q_d/Q=1/2$.

The propagation of capsules through the T-junction sorting device is shown in figure \ref{fig:eg-PFF-TJ}, with a sequence of snapshots taken at variable time intervals. The final position of each capsule in the diffuser is quantified by the offset of the capsule centroid from the centreline of the diffuser, $\lambda$. Each value of $\lambda$ was obtained by averaging $y$-positions of the centroid for $x= 24.5$~mm, where the origin of the $x$ axis is chosen as the left wall of the main channel (see figure \ref{fig:TrajAndWidth-PinTJ}). Figures \ref{fig:eg-PFF-TJ}(a,b) show a comparison between the trajectories of the same capsule (P4) for two values of the flow rate $Q$. It indicates a significant reduction of $\lambda$ with increasing flow rate, which is associated with enhanced capsule deformation in the T-junction and the turning of the capsule into the diffuser closer to the outlet of the main channel. Figures \ref{fig:eg-PFF-TJ}(a,c) show two different capsules (P4, P1) subject to the same flow rate. The softer capsule (P1) has approximately the same value of $Ca$ as capsule (P4) in figure \ref{fig:eg-PFF-TJ}(b). As a result, it exhibits a deformation and trajectory in the T-junction very similar to that shown in figure \ref{fig:eg-PFF-TJ}(b), resulting in a similar reduced value of $\lambda$. This suggests that the T-junction sorting device can separate capsules according to stiffness and that the process is primarily governed by $Ca$ for capsules of fixed size. 

Figure \ref{fig:TrajAndWidth-PinTJ} shows the trajectories of the centroids of three capsules (P1, P3, P4) spanning a ninefold variation in stiffness, at a constant value of the flow rate. The data shows that the capsule centroids follow distinct paths in the diffuser, resulting in monototonically increasing values of $\lambda$ with increasing stiffness. This separation is initiated inside the T-junction where the distance from the end-wall at which the centroids turn into the diffuser decreases monotonically with increasing capsule stiffness (see inset of figure \ref{fig:TrajAndWidth-PinTJ}).  This initial separation is then significantly enhanced as the capsules are entrained into the diffuser where streamlines diverge.

\begin{figure}
	\centering
	\includegraphics[width=1.0\textwidth]{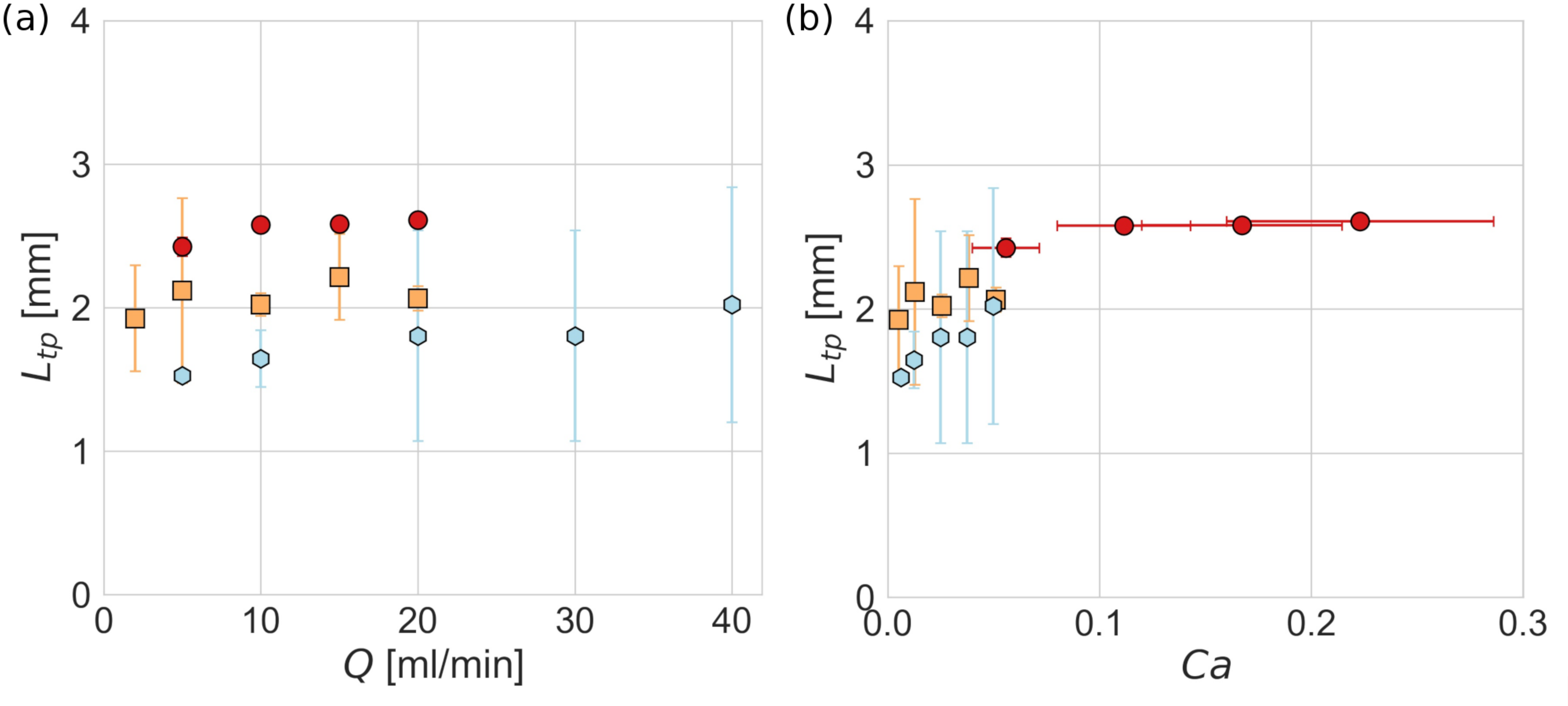}
	\caption{Turning point of capsules in the sorting device operated in T-junction mode as a function of  (a) flow rate $Q$ and (b) capillary number $Ca$ based on the flow rate $Q$ in the main channel. Red circles: capsule P1; orange squares: P3; light blue hexagons: P4. The ratio of flow rates is $Q_d/Q=1/2$.
	}
	\label{fig:Turning-Pinching}
\end{figure}
\begin{figure}
	\centering
	\includegraphics[width=1.0\textwidth]{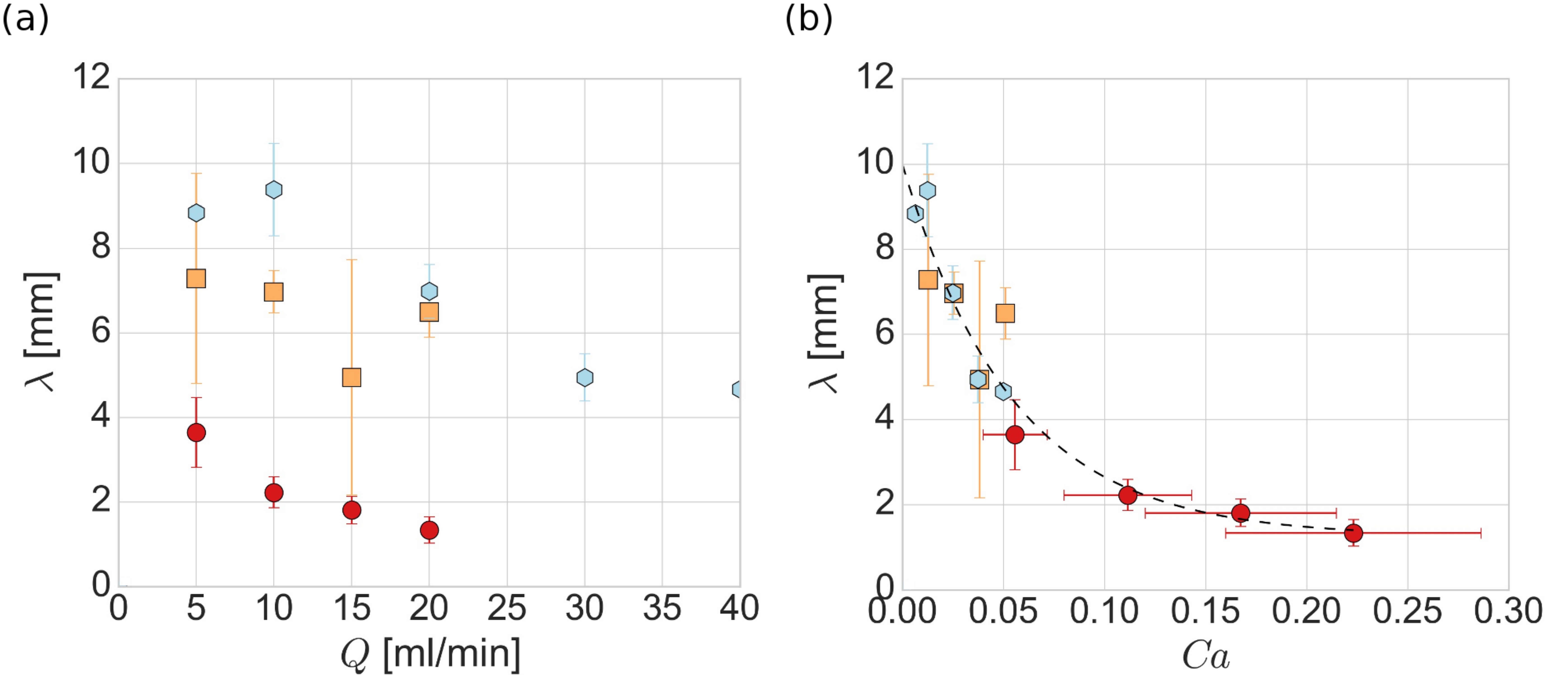}
	\caption{ Variation of the final position of the capsule centroid relative to the channel centreline, $\lambda$, with (a) flow rate $Q$, (b) capillary number $Ca$, for $Q_d/Q = 1/2$.  
		Red circles: capsule P1; orange squares: P2; light blue pentagons:	P4. 
		The dashed black line is an exponential fit to the data to guide the eye.
	}\label{fig:SortingTJ}
\end{figure}

In figure \ref{fig:SortingTJ}, we quantify the distance from the end-wall of the T-junction at which the capsules are entrained into the diffuser, $L_{tp}$ (see figure \ref{fig:TpTJ} for definition), as a function of $Q$ and $Ca$. At constant flow rate, $L_{tp}$ increases monotonically with stiffness as previously illustrated in figure \ref{fig:TrajAndWidth-PinTJ}, and collapses approximately onto a monotonically growing master curve as a function of $Ca$. This behaviour contrasts with the results in the T-junction flow device, where a small reduction in $L_{tp}$ was observed for increasing $Ca$ (see figure \ref{fig:TpTJ}). We believe that this trend reversal is due to the fact that in the sorting device considered in the current section the main channel
only has half the width of that in the T-junction setup
studied in \S \ref{subSec:TJMode}. Figure \ref{fig:eg-PFF-TJ} shows that the resulting much
stronger confinement causes the capsule to be more elongated
when it enters the T-junction. As a result, a significant fraction of its
length is still confined by the narrow main channel when its front
begins to bend towards the diffuser. Stiffer capsules offer more
resistance to this deformation  and therefore turn slightly
later than their softer counterparts (see figure \ref{fig:TrajAndWidth-PinTJ}). The resulting
slight difference in the trajectory of the centroid through the T-junction
shown in the inset in figure \ref{fig:TrajAndWidth-PinTJ} is then enhanced in the diffuser, leading
to a noticeable separation of capsules of different stiffness.


The variation of the final position of the capsule centroid relative to the centreline of the diffuser, $\lambda$, is shown in Figure \ref{fig:SortingTJ}(a, b) as a function of flow rate and capillary number, respectively. $\lambda$ decreases with increasing flow rate for each capsule, which is consistent with the increasing distance of the turning point of the capsule from the end wall of the junction (see Figure \ref{fig:Turning-Pinching}). Again, the data collapses onto a master curve when plotted as a function of $Ca$ as predicted from figure \ref{fig:eg-PFF-TJ}. A fit to the data yields minimum and maximum values of $\lambda$ of $\lambda_{min}=  1.6 \pm 0.6$~mm for large values of $Ca$ and $\lambda_{max}=9.7 \pm 1$~mm for rigid particles ($Ca \rightarrow 0$).

Capsules P4 and P1, which differ by a factor 9 in stiffness are clearly differentiated for all flow rates. Their average centroid positions are separated by approximately 5 mm, which is sufficient for sorting into distinct channels at the end of the diffuser given that the capsule diameter is in the range 3.8 -- 3.9 mm.
The stiffness factor between P4 and P3 is only 1.5, and they reach the end of the diffuser with a separation of 1.5 mm for a flow rate of $Q=10$~ml/min. They are clearly separated in the diffuser and a suitable design of the capture channels would also allow their sorting in practice.

\subsection{PFF mode of operation}
\label{sec:pinchingMode}

\begin{figure}
	\centering
	\includegraphics[width=1.0\textwidth]{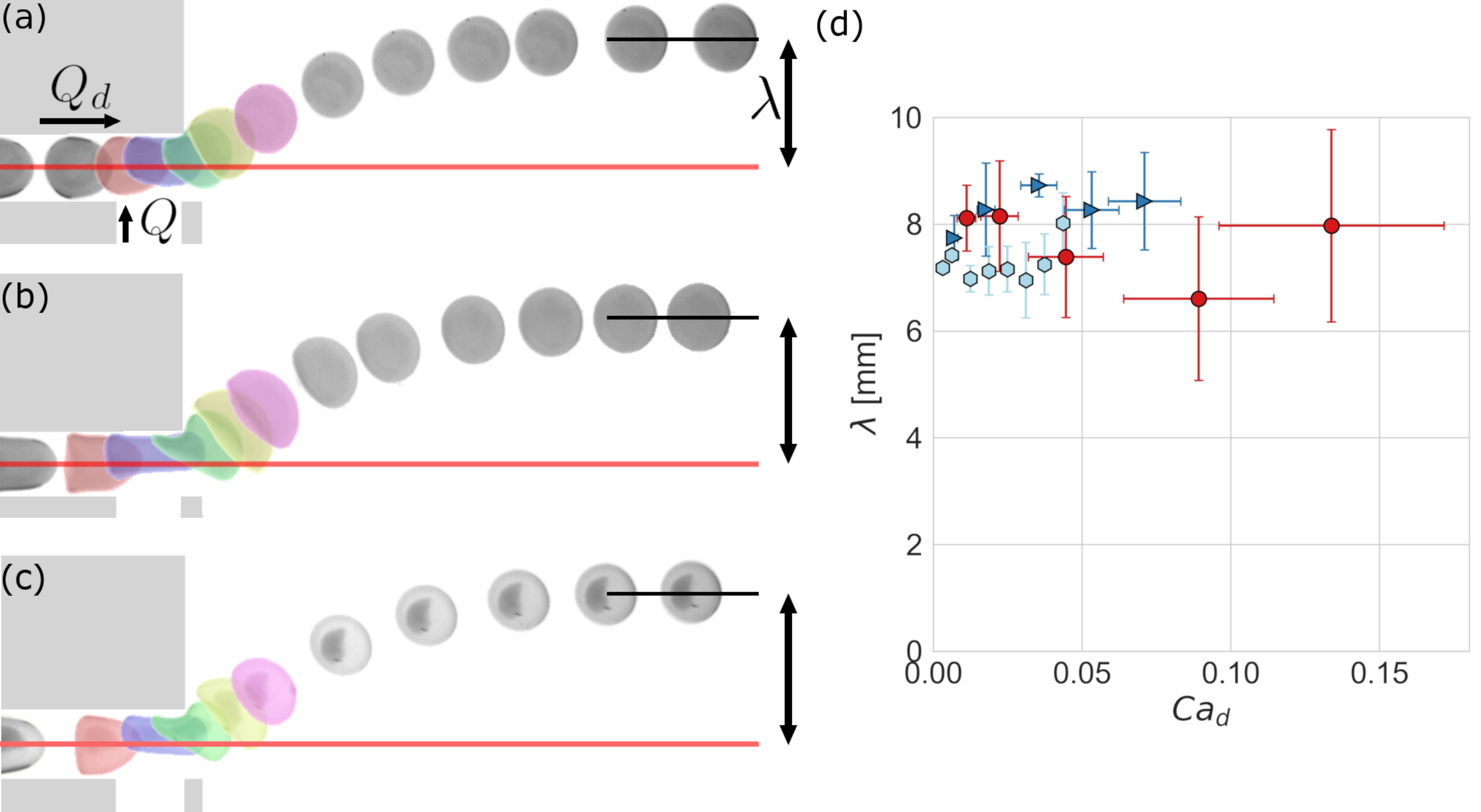}
	\caption{Sequences of images illustrating the trajectory of a capsules through the sorting device in the pinching mode. The centreline of the daughter channels is shown in red.
		The capillary number $Ca_d$ is based on the flow rate $Q_d$ and cross-section of the narrow daughter channel.
		(a)  P4, $Q =  5$~ml/min, $Q_d/Q=1/2$
		($Ca_d= (2.9 \pm 0.08) \times 10^{-3}$).
		(b) P4, $Q  =  75$~ml/min, $Q_d/Q=1/2$ 
		($Ca_d= (4.4 \pm 0.12) \times 10^{-2}$).
		(c) P2, $Q  =  5$~ml/min, $Q_d/Q=1/2$ 
		($Ca_d= (3.6 \pm 0.61) \times 10^{-2}$). 
		(d) Final displacement of the capsule centroid from the channel centreline as a function of $Ca_d$.  Red circles: capsule P1; dark blue triangles: P2; light blue pentagons: P4. 
	}\label{fig:eg-PFF-pinch}
\end{figure}

The sorting device employed in \S \ref{subSec:TJMode} can alternatively be operated as a pinched flow fractionation device (PFF), which is commonly utilised to separate rigid particles of different sizes \citep{YamadaEtAl04, CupelliEtAl13}. We explore whether PFF can be used to separate deformable particles according to their stiffness. For this purpose, capsules were propagated into the T-junction through the left hand-side daughter channel by imposing different values of the flow rate $Q_d$. A second flow rate $Q$ was imposed in the main channel to generate a jet orthogonal to the main stream that `pinches' the capsule against the end wall of the junction. All experiments were performed for $Q_d/Q=1/2$.

Figure \ref{fig:eg-PFF-pinch}(a,b,c) shows three sequences of images illustrating the trajectories of capsules through the sorting device operated in the pinched flow fractionation mode (PFF). Here, the capsules are propagated towards the diffuser through the left hand-side daughter channel with flow rate $Q_d$, while the pinching flow is imposed through the main channel with flow rate $Q$. The ratio of the imposed flow rates is kept constant at $Q_d/Q=1/2$ in all experiments. In figure \ref{fig:eg-PFF-pinch}(a,b), the same capsule (P4) propagates through the device for two different values of the flow rate. The different flow rates affect the propagation in the inlet channel, where the capsule is weakly deformed at low flow rate and buckled into a parachute shape at high flow rate. However, the stronger pinching flow in figure \ref{fig:eg-PFF-pinch}(b) does not result in significantly increased values of $\lambda$, despite larger deformation of the capsule than in figure \ref{fig:eg-PFF-pinch}(a). In Figure \ref{fig:eg-PFF-pinch}(c) a capsule softer by a factor of almost six (P2), which is subject to the same flow rate as capsule (P4) in figure \ref{fig:eg-PFF-pinch}(a), follows a similar trajectory so that its final offset does not differ significantly from those in figure \ref{fig:eg-PFF-pinch}(a,b). This means that the sorting device operated in PFF mode is not effective at separating capsules according to stiffness.
This result is confirmed in figure \ref{fig:eg-PFF-pinch}(d), where $\lambda$ measured for three capsules of the same size but of different stiffness (P1, P2, P4) is approximately constant as a function of $Ca_d$ (capillary number based on the flow rate $Q_d$). This is because the displacement of the capsule centroid towards the channel wall opposite the main channel outlet is not sufficient to allow them to diverge in the diffuser. This indicates that pinched flow fractionation cannot be used to sort capsules according to their stiffness with the current setup.

\section{Discussion and conclusion}

We have characterised the transport of centred capsules of fixed size through a symmetric T-junction as a function of their stiffness and of the imposed flow rate, with extensive experiments on a small number of carefully selected capsules. Each capsule was subjected to large deformations in the T-junction in up to 90 consecutive experiments without developing any measurable permanent deformation, thus confirming that plastic deformation was negligible in these experiments. We find that the motion and deformation of the capsules through the T-junction characterised in terms of their fractional speed in the straight channels, their deformation and turning point in the T-junction, as well as the distance over which they relax back to a steady configuration in the daughter channels, can be described by a capillary number based on the elastic force required to statically compress the capsule between parallel plates to 50\% of its undeformed diameter. This global measure of elastic resistance includes the effect of non-linear elastic deformation, pre-inflation of the capsule and thickness of the encapsulating membrane. 

We have exploited the large $Ca$-dependent deformation of capsules in the T-junction to devise a novel approach to measuring relative capsule stiffness, which we find particularly suited to differentiate between our softest capsules. This increased resolution means that the method complements compression measurements commonly used to measure the stiffness of millimetric capsules \citep{CarinEtAl03, Risso2004, RachnikEtAl06}, where the softest capsules are often difficult to distinguish. This is because of the technical challenges associated with the measurement of forces of sub-Newton magnitude and the manipulation of these fragile objects. In-flow measurements are routinely performed on micrometric capsules suspended in a carrier liquid within a microchannel. Their shear modulus is typically determined by comparison with numerical simulations of the deformation in flow along a straight section of tube of capsules based on a membrane model  \citep{HurEtAl11, Hu2013}. However, the range of deformations achievable in a straight section of tube is narrower than that in a T-junction, and thus the accuracy of the method could be enhanced by focusing on the flow through a T-junction. Based on their numerical simulations, \cite{Koolivand2017} recently proposed a method involving the trapping of capsules in a T-junction which relied on time-consuming user input to trap a single capsule through skilled micro-manipulation. In contrast, our strategy of measuring maximum length in flow within the T-junction does not utilise trapping and thus, this method could be automated to reliably test populations of capsules of known size. 

We also exploited the $Ca$-dependent turning point of the capsules in the T-junction to sort capsules according to their stiffness. By broadening the channel at the exit of the T-junction, we increased the range of turning positions for our range of $Ca$ and by transporting them along diverging streamlines in this diffuser, we further enhanced the separation of capsules exiting the T-junction for different values of $Ca$. By comparison with the half-cylinder sorting device proposed by \cite{Zhu2014} and recently investigated experimentally by \cite{Haener2017}, our device achieves larger separations by a factor of approximately two, for similar diffuser dimensions. In fact, our sorting device was originally designed to explore capsule separation in both pinching flow and T-junction modes, but we found that only the T-junction mode could robustly sort capsules according to their stiffness. Our prototype device is sufficient to deliver a proof-of-concept, but there remains considerable scope for its optimisation. A second-generation T-junction device would encompass symmetric daughter channels both acting as diffusers, so that the device could be operated with a single syringe pump infusing liquid into the main channel. Moreover, the dimensions of the main channel, T-junction chamber and diffusers could be optimised to maximise the range of turning points positions of the capsules, and capsule offsets, respectively. 


\appendix
\section{Capsule Manufacture}\label{appA}

An aqueous solution containing protein and gelling agent was prepared by dissolving 1\% of sodium-alginate (Sigma), 2\% of propylene glycol alginate (PGA,  Brewpaks) and 8\% of ovalbumin (62-88\%, agarose gel electrophoresis, Sigma), by weight, in distilled water. The mixture was agitated with a magnetic stirrer for an hour, before storing it at $4^\circ$C for 12 hours. The foam formed on the surface was then
removed and any undissolved reactant broken up with a spatula before agitating the mixture again and refrigerating it until use.

This solution was then added dropwise from a syringe needle to a 10\% solution of calcium chloride (CaCl$_2$, anhydrous granular, size $ \le 7.0$~mm, $\ge 93\%$, Sigma) and the reaction of the drops with the sodium alginate solidified them into gel beads. The height at which the needle was positioned above the bath was continually adjusted to maximise the sphericity of the beads. The calcium carbonate solution was agitated with a magnetic stirrer for 10 minutes to allow the beads to gel fully before discarding it and rinsing the beads three times in an 11 g/l NaCl aqueous solution (saline). The beads were then immersed in an aqueous alkaline solution (NaOH, $10^{-4}$ M to $10^{-3}$ M, Sigma) and stirred for 5 minutes. This resulted in a transacylation reaction on the surface of the beads between ester groups of the PGA and unionised amino groups in the ovalbumin, which led to the creation of amide bonds so that each bead was encapsulated by a cross-linked membrane. The concentration of sodium hydroxide determined the degree of cross-linking of the membrane and its thickness, and hence the stiffness of the membrane. The alkaline solution was then neutralised with hydrochloric acid (Sigma) of the same molarity and stirred again for 5 minutes. The beads were rinsed three times with saline before placing them in a solution of sodium citrate ($\ge 99\%$, Sigma) to re-liquefy the gel core. The resulting capsules were rinsed again three times before storing them in saline. Solid beads were made in a similar manner to capsules. In order to maximise their stiffness, the concentration of the alkaline solution was increased to $0.1$M, and the beads were not exposed to sodium citrate.

\bibliographystyle{jfm}
\bibliography{Capsules_Anne}

\end{document}